\documentclass[aps,twocolumn,superscriptaddress,nofootinbib,showpacs,amsmath,amssymb,superscriptaddress]{revtex4}
\usepackage{graphicx}
\usepackage{colordvi}
\usepackage[dvips]{color}
\usepackage{amsmath}
\usepackage{bm}
\newcommand{\bea}{\begin{eqnarray}}
\newcommand{\eea}{\end{eqnarray}}
\newcommand{\be}{\begin{equation}}
\newcommand{\ee}{\end{equation}}
\def\gtwid{\mathrel{\raise.3ex\hbox{$>$\kern-.75em\lower1ex\hbox{$\sim$}}}}
\def\ltwid{\mathrel{\raise.3ex\hbox{$<$\kern-.75em\lower1ex\hbox{$\sim$}}}}

\begin{document}
\title {Local quasiparticle lifetimes in a $d$-wave superconductor}

\author{S. Graser}
\email{graser@phys.ufl.edu} \affiliation{Physics Department,
University of Florida, Gainesville, FL 32611 USA}

\author{P.J.~Hirschfeld}
\email{pjh@phys.ufl.edu}
\affiliation{Physics Department,
University of Florida, Gainesville, FL 32611 USA}

\author{D.J.~Scalapino}
\email{djs@physics.ucsb.edu} \affiliation{Department of Physics,
University of California, Santa Barbara, CA 93106-9530 USA}

\date{\today}
\begin{abstract}
Scanning tunnelling spectroscopy (STS) measurements find that the
surface of Bi-2212 is characterized by nanoscale sized regions,
``gap patches," which have different magnitudes for the $d$-wave
energy gap $\Delta_0({\bf r})$. Recent studies have shown that the
tunnelling conductance can be fit using a BCS-type density of
states for a $d$-wave superconductor with a {\it local}
quasiparticle scattering rate. The fit is made with a scattering
rate which varies linearly with energy and has a slope
$\alpha({\bf r})$ that is positively correlated with the local value
of the gap.  We consider first the question of what is actually
measured in such an experiment. To this end we revisit a model of
quasiparticle scattering by impurities and spin fluctuations which
was previously used to describe the lifetimes of nodal
quasiparticles measured by angle-resolved photoemission (ARPES).
We argue that the broadening of the local density of states is
determined, except in the case of localized impurity bound states,
by the imaginary part of the self-energy of the system averaged
over a small region. The size of this region is set by a mean
free path which depends upon the energy. At low energies, this
region is found to be significantly larger than a gap ``patch", so
that the density of states measured by STS is homogeneous in this
energy range. At higher energies where the mean free path is
comparable with the patch size, the density of states is
inhomogeneous. We show that a local self-energy in the
impurity-plus-spin fluctuation model, while not strictly linear,
yields a local density of states (LDOS) nearly identical to the
full theory, and argue that it is consistent with the STS data as
well as the phenomenological linear scattering rate extracted from
experiment. We also explore the qualitative consequences of this
phenomenology for the spectral widths observed in ARPES and
predict the existence of Fermi arcs in the superconducting state.
\end{abstract}


\maketitle

\section{Introduction}
\label{sec:intro}

For many years, it has been recognized that the properties of the
quasiparticle states near the nodes of the $d$-wave
superconducting gap in the cuprates were quite different from
those near the gap maxima or antinodes. Transport estimates of
nodal mean free paths at low temperatures range from tens of
$\mu$m in the clean stoichiometric YBCO
compounds\cite{RHarris:2006} to hundreds of \AA~in dirtier
materials like BSCCO\cite{SOzcan:2006}. ARPES measurements on
BSCCO, with much cruder energy resolution, have reported residual
nodal widths of the momentum distribution curve (MDC) of about
10meV\cite{TYamasaki:2006}, which with Fermi velocity values of
$10^{7}$ cm/s imply much shorter mean free paths, of order 30 \AA.
While this discrepancy is not currently understood, both transport
and ARPES {\it nodal} mean free paths are significantly longer
than those extracted from ARPES for {\it antinodal}
quasiparticles. These scattering rates are of order 25-40meV at
low $T$ in the optimally doped superconducting
state\cite{antinodeARPES}, corresponding naively to mean free
paths of order a unit cell size. One explanation for this
``dichotomy" between nodal and antinodal states that has been
proposed is the existence of intense magnetic scattering near the
antinodal point in the underdoped materials due to the proximity
of the wave vector $\bf Q$ connecting the antinodal points by an
antiferromagnetic nesting vector.  This scattering has also
recently been related by several authors to the properties of the
``arcs" of coherent Fermi surface in the pseudogap
state\cite{JGStorey:2007,AVChubukov:2007,MRNorman:2007}.

Recently, a new perspective has been brought to bear on the
properties of these quasiparticle states by STS. In
Ref.~\cite{JAlldredge:2007}, Alldredge and collaborators succeeded
in fitting STS data on Bi-2212 at various dopings to an extremely
simple form, that of the local density of states (LDOS) of a
$d$-wave superconductor with a single position and frequency
dependent broadening function or scattering rate
$\Gamma({\bf r},\omega)$, \bea N({\bf r},\omega) = \left\langle
\frac{\omega +i\Gamma({\bf r},\omega)}{(\omega +
i\Gamma({\bf r},\omega))^2
 - \Delta_1^2({\bf r}) \phi_d({\bf k})^2} \right\rangle. \label{dwaveLDOS}
\eea Here \bea \Gamma({\bf r},\omega) = \Gamma_1({\bf r})+
\Gamma_2({\bf r},\omega) = \Gamma_1({\bf r}) + \alpha({\bf r}) \omega,
\label{proportion} \eea with $\phi_d({\bf k})=(\cos k_x -\cos k_y)/2$
and  $\alpha({\bf r})$  a constant of proportionality which was found
to vary from point to point over the sample surface, and to be
strongly and positively correlated with the size of the local gap
parameter $\Delta_1({\bf r})$. The frequency independent term
$\Gamma_1$ simulates the effect of near unitarity limit in-plane
scatterers, which are observed as ``native defect" resonances in
conductance maps of the Bi-2212 surface.   Values of $\Gamma$
extracted from this fit procedure tended to be much smaller than
quasiparticle widths determined by ARPES, ranging at energy
$\omega = \langle \Delta_1 \rangle$ from about 2 meV for an
optimally doped sample to about 20 meV for a strongly underdoped
sample.   These results, and the apparent contradiction with
ARPES, lead us to consider the following questions:
\begin{itemize}
\item What sort of scattering rate is actually being measured in
the STS measurement?

\item When is it possible to think of a local probe measuring a
scattering rate averaged over a region or over disorder, and what
length scales define such a region?

\item How can the spatially modulated $\Gamma_2$ at {\it low}
energies implied by Eq.~(\ref{proportion}) be reconciled with the
observed {\it homogeneity} of the low-energy quasiparticle states
seen in STS?

\item Are the results of Ref.~\cite{JAlldredge:2007} possibly
consistent with scattering rates other than pure linear in
$\omega$, derivable from microscopic models?

\item How are the small values of the STS-determined scattering
rates to be reconciled with ARPES?

\end{itemize}

In the following we attempt to answer these questions within a
phenomenological framework which assumes as its starting point
that there are nanoscale energy gap inhomogeneities.  Various
suggestions regarding the origin of this ``gap patch"
inhomogeneity have been
proposed\cite{IMartin:2001,ZWang:2002,SAKivelson:2003,TSNunner:2005,JXZhu:2006,ACFang:2006,AVBalatsky:2006}.
Here we will simply assume that for $T\ll T_c$, the
superconducting gap has such nanoscale patch inhomogeneities. We
will then proceed in several stages.  In Section~\ref{sec:general}
we discuss the local density of states of a disordered system, and
the conditions under which a system may be considered to be
locally self-averaging.  We then extend these considerations to
inelastic scattering, and discuss what may be deduced from the
qualitative aspects of the STS results. In
Section~\ref{sec:scattering} we introduce a model for scattering
from impurities, including native defects in the CuO$_2$ planes,
and dopant disorder away from the planes.  To the
disorder-averaged impurity self-energy we then add the self-energy
due to scattering from spin fluctuations, and discuss the
anisotropy of the total scattering in momentum space.  We then
calculate the LDOS from the local Green's function in
Section~\ref{sec:localG}, and compare the results with those
computed from a model ``local self-energy".  We show that the
spectrum may equally well be fit by an assumed linear scattering
rate, as proposed in Ref.~\cite{JAlldredge:2007}. Finally, we
discuss the extension of the results obtained for the homogeneous
$d$-wave superconductor to the inhomogeneous case.  We discuss how
the spatially homogenous low-energy conductance seen in the STS
experiments for $\omega\ltwid\Delta_0$ and the heterogeneous
behavior for $\omega\gtwid\Delta_0$ may be understood within this
framework.  In addition, we point out the implications of
Ref.~\cite{JAlldredge:2007} for the interpretation of ARPES
spectra in the superconducting state.

\section{General considerations}
\label{sec:general}  We begin by assuming that STS probes the
LDOS, i.e. the imaginary part of the  local Green's function
$G({\bf r},{\bf r})$; that is, we assume that the tunnelling matrix
elements are constant. For the moment we focus on disorder in the
normal metallic state, and neglect interactions. In this case one
should in principle calculate $G$ for a given configuration of
impurities. There is no notion of a spatially or disorder averaged
self-energy which need appear, if one can calculate $G({\bf r},{\bf r})$
for a system with disorder.  Indeed, the exact local Green's
function may be written \bea G({\bf r},{\bf r};\omega) = \sum_n
\frac{|\psi_n({\bf r})|^2}{\omega - E_n +i0^+}, \eea where
$\psi_n({\bf r})$ are the exact eigenfunctions of the many-electron
system in the presence of a given random disorder configuration.
Such a calculation would include all interference processes from
these many impurities.

Note that even if we are interested exclusively in the local
Green's function, it is far from obvious that one may speak of a
local self-energy or scattering rate, as proposed by the STS
experiments. This is because the self-energy is formally a
nonlocal quantity as is seen from the Dyson equation for
$G({\bf r},{\bf r})$: \bea G({\bf r},{\bf r}) = G_0(0) +
G_0({\bf r}-{\bf r}')\Sigma({\bf r}',{\bf r}'')G({\bf r}'',{\bf r}), \eea where $G_0$ is
the Green's function of the homogeneous system. $\Sigma$ may be
treated as local, $\Sigma({\bf r},{\bf r}')\sim
\Sigma({\bf r})\delta({\bf r}-{\bf r}')$ only if $\Sigma({\bf r},{\bf r}')$ decays
sufficiently rapidly.

Under some circumstances is it appropriate to approximate the
many-impurity Green's function by a homogeneous disorder-averaged
Green's function $\bar G$. In a dirty metal, the wave functions
$\psi({\bf r})$ decay on a length scale of the mean free path $\ell$
of the system, as do the disorder averaged Green's functions and
self-energies. This will then be a good description of the system
{\it provided} there are no bound state wave functions present
where electrons are confined to within a radius $\ell_b$
substantially smaller than $\ell$. Near the bound state energy
$\bar G({\bf r},{\bf r};\omega)$ will provide a poor description of the
spatial and spectral properties of this state.

Similar considerations apply to quasiparticles in superconductors.
In general, a local measurement by STS will probe wavefunctions
spread out over a mean free path.  However, one should be wary in
a highly anisotropic system like a $d$-wave superconductor of
assuming the relevance of mean free paths extracted from other
measurements. For example, transport measurements at low $T$ yield
the mean free paths of nodal quasiparticles only, and can be very
long, whereas an STS experiment probing the system locally at bias
$\omega$ will involve quasiparticles with a momentum spread
$\omega/v_F$ and these may well have shorter mean free paths.  We
will see that results of the STS measurements may be used to
determine the typical mean free path of the excitation being
measured. In general, however, the system may indeed be taken to
be self-averaging and a treatment in terms of a disorder-averaged
$\bar G$ should be adequate.

An exception occurs again in the case of impurity bound states,
known to occur in the cuprate materials for Cu substituents in the
CuO$_2$ plane. As observed by STM, in these cases a very narrow
spectral resonance is found at frequency $\Omega_0$ on the
impurity site and near neighbors.   The naive calculation of the
size of these states at resonance yields \bea \ell_b =
\frac{\xi_0}{\sqrt{1-{\Omega_0^2\over \Delta_0^2}}}, \eea where
$\Delta_0$ is the gap maximum and $\xi_0=v_F/(\pi\Delta_0)$,
except in the nodal directions, where $\psi_b\sim 1/r$ and $\xi_0$
diverges.  Note that the electron unbinds when the resonance moves
into the continuum $\Omega_0>\Delta_0$, corresponding to weak
impurity potential. More detailed analysis shows that the length
scale $\ell_b$ given above represents the asymptotic long-range
decay of the wavefunctions in non-nodal directions; most of the
decay away from the impurity site actually takes place over an
atomic distance. In any case, for in-plane impurities, $l_b \ll
l$, so they should be (and are) imaged individually by STS.  On
the other hand, the out-of-plane dopant atoms in the BSCCO-2212
material have weak potentials and no resonance, corresponding to
an essentially infinite $l_b$, i.e. a disorder averaged
calculation of the local Green's function is again appropriate. We
conclude that a disorder-averaged $\bar G$ is a good
representation of the local exact $G({\bf r},{\bf r})$ over a length scale
of size $\ell$ except within a distance $\ell_b$ of strong
impurities. Therefore the notion of a local self-energy or
scattering rate will also have meaning, except for a small set of
exceptional points.

\section{Model for scattering}
\label{sec:scattering}

The purpose of this section is to present the results of a
scattering-rate  analysis of spin fluctuations and impurities
which will be used to determine the local $\bar G$. We assume that
a self-averaged description is appropriate, and investigate, in
the BSCCO system, what the sources of scattering are, and how they
enter the local $\bar G$. We can then ask if a description in
terms of a local self-energy or scattering rate is sensible or
not, and whether or not a fit to the LDOS obtained from $\bar G$
can also be obtained with a linear scattering rate as proposed in
Ref.~\cite{JAlldredge:2007}. This will give us information as to
the robustness of such a fit.

\subsection{Elastic potential scattering}\label{subsec:elastic}

\begin{figure}
\begin{center}
\includegraphics[width=.45\columnwidth]{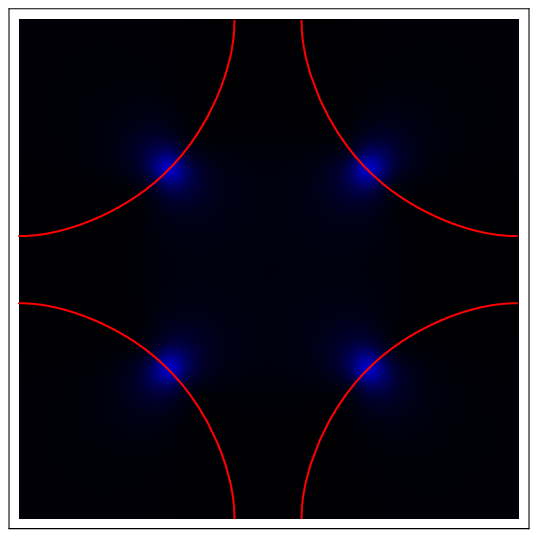}
\includegraphics[width=.45\columnwidth]{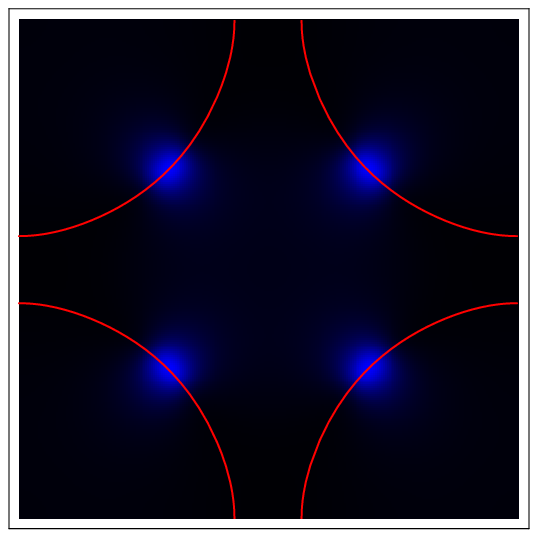}
\includegraphics[width=.45\columnwidth]{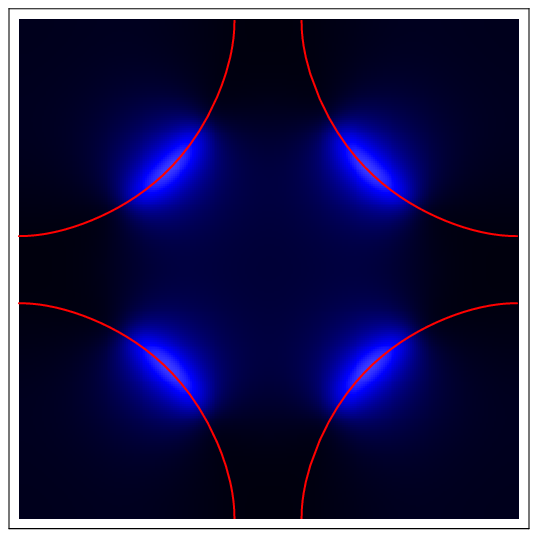}
\includegraphics[width=.45\columnwidth]{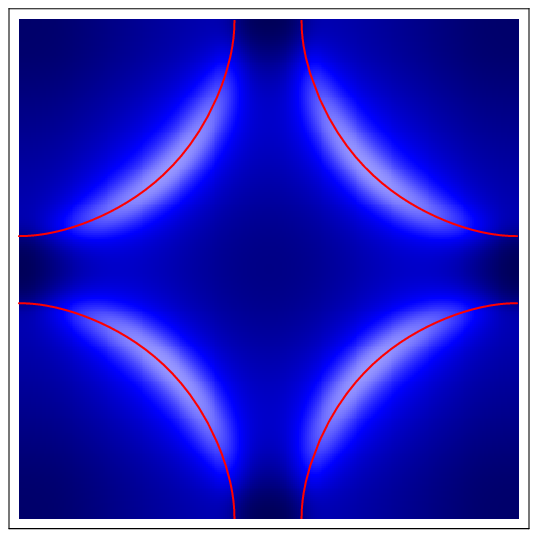}
\end{center}
\caption{(Color online) The first two terms of Eq.
(\ref{Eq:effective_scattrate}) ($\tau_0$ and the $\tau_1$
contributions) to the elastic scattering rate as a function of
$k_x$ and $k_y$ in the first Brillouin zone for different
quasiparticle energies $\omega$. From left to right and top to
bottom $\omega$ is chosen to be $\omega = 0.1 \Delta_0$, $\omega =
0.2 \Delta_0$, $\omega = 0.5 \Delta_0$ and $\omega = \Delta_0$.
The red line shows the Fermi surface obtained from the single
$t-t'$ band described in the text. The inverse range of the
impurity potential is $\kappa = 0.2$, corresponding to strong
forward scattering.  The color scale is the same for all four
plots.} \label{fig:ElasticBZ02}
\end{figure}

The elastic scattering in BSCCO can arise from impurities and
disorder inside the conducting copper oxide planes (in-plane
impurities) or due to impurities in the neighboring metal oxide
planes (out-of-plane impurities). While the in-plane impurities
are well characterized by a point-like isotropic scattering
potential, the out-of-plane impurities are better described by a
long range e.g. Yukawa type impurity potential leading to a more
or less pronounced forward
scattering~\cite{EAbrahams:2000,TDahm:2005a,TDahm:2005b}. Since
the out-of-plane impurity scattering is relatively weak it can be
treated within the self-consistent Born approximation. The elastic
self-energy is then given by \be {\underline\Sigma}_{\rm el}
({\bf k},\omega) = n_i \sum_{k'} |V({\bf k},{\bf k}')|^2 \tau_3 {\underline
G}({\bf k}',\omega) \tau_3 \ee where $\tau_\alpha$ are the Pauli
matrices in particle hole space, $n_i$ is the impurity
concentration and $V({\bf k},{\bf k}')$ characterizes the momentum dependent
impurity potential. Within our calculations we will model it by a
screened exponential falloff of the form \be
|V({\bf k},{\bf k}')|^2=\frac{|V_0|^2}{|{\bf k}-{\bf k}'|^2+\kappa^2} \ee where
$\kappa^{-1}$ characterizes the range of the impurity potential.
For $\kappa\rightarrow\infty$ the case of isotropic scattering is
recovered. The full Green's function $G({\bf k},\omega)$ can be
calculated from the Dyson equation \be {\underline
G}^{-1}({\bf k},\omega) = {\underline G}_0^{-1}({\bf k},\omega) -
{\underline \Sigma}_{\rm el} ({\bf k},\omega) \ee where
$G_0({\bf k},\omega)$ describes the pure superconducting state without
scattering. Decomposing the self energy into its different Nambu
components leads us to the following three equations which must be
solved self-consistently \be \Sigma_{\rm el,0}({\bf k},\omega)=n_i
\sum_{k'}|V({\bf k},{\bf k}')|^2
\frac{\tilde{\omega}}{\tilde{\omega}^2-\tilde{\epsilon}_{{\bf k}'}^2-\tilde{\Delta}_{{\bf k}'}^2},
\ee \be \Sigma_{\rm el,1}({\bf k},\omega)= - n_i\sum_{k'}|V({\bf k},{\bf k}')|^2
\frac{\tilde{\Delta}_{{\bf k}'}}{\tilde{\omega}^2-\tilde{\epsilon}_{{\bf k}'}^2-\tilde{\Delta}_{{\bf k}'}^2}
\ee and \be \Sigma_{\rm el,3}({\bf k},\omega)=n_i\sum_{k'}|V({\bf k},{\bf k}')|^2
\frac{\tilde{\epsilon}_{{\bf k}'}}{\tilde{\omega}^2-\tilde{\epsilon}_{{\bf k}'}^2-\tilde{\Delta}_{{\bf k}'}^2}
\ee Here $\tilde{\omega}=\omega-\Sigma_{\rm el,0}({\bf k},\omega)$,
$\tilde{\epsilon}_{\bf k}=\epsilon_{\bf k}+\Sigma_{\rm el,3}({\bf k},\omega)$ and
$\tilde{\Delta}_{\bf k}=\Delta_{\bf k}+\Sigma_{\rm el,1}({\bf k},\omega)$ are the
renormalized quasiparticle energy, band structure and pairing
potential respectively. For our numerical calculations, we will
use a simple tight-binding parameterization of the band structure
including nearest and next nearest neighbor hopping \be
\varepsilon_{\bf k}=-2t\, (\cos k_x+\cos k_y) - 4t' \cos k_x \cos k_y -
\mu \label{band}\ee with a $d$-wave order parameter \be \Delta_{\bf k}
= \frac{\Delta_0}{2}\ (\cos k_x-\cos k_y)\, . \ee To model the
Fermi surface in the cuprates we use $t' = -0.35 t$ and $\mu=-1.1
t$, with energy measured in units of nearest neighbor hopping $t$.
The temperature dependence of the gap is parameterized by the
simple form \be \Delta_0(T) = \Delta_0 \tanh
\left(\alpha\sqrt{\frac{T_c}{T}-1}\right) \ee with $\alpha=3$,
$2\Delta_0/k\, T_c=6$, and $\Delta_0=0.2 t$.

The elastic scattering rate, broadening the quasiparticle state
with energy $\omega$, is determined by the denominator of the full
Green's function and can be calculated approximately from the pole
of $G$ as \bea \Gamma_{\rm el}({\bf k}, \omega) & = & - {\rm Im} \left(
\Sigma_{\rm el,0}({\bf k},\omega)  + \frac{\Delta_{\bf k}}{\omega}
\Sigma_{\rm
el,1}({\bf k},\omega) \right. \nonumber \\
& & \left. +  \frac{\epsilon_{\bf k}}{\omega}  \Sigma_{\rm
el,3}({\bf k},\omega)   \right) \label{Eq:effective_scattrate}\eea For
quasiparticles at the Fermi surface only the imaginary parts of
$\Sigma_{\rm el,0}$ and $\Sigma_{\rm el,1}$ contribute to the
elastic scattering rate. To compare our results for different
impurity ranges, we have normalized the scattering rates in a way
that the (normal conducting) scattering rate for $T_c$ taken at
the nodal point of the Fermi surface ${\bf k}_N$ equals $0.1 \Delta_0$.
All calculations for the elastic as well as for the inelastic
scattering rate have been performed for a low temperature $T= 0.1
T_c$ unless otherwise stated.

\begin{figure}
\begin{center}
\includegraphics[width=.45\columnwidth]{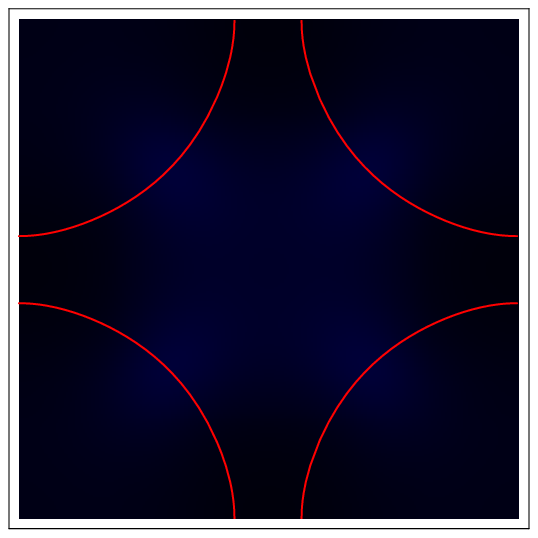}
\includegraphics[width=.45\columnwidth]{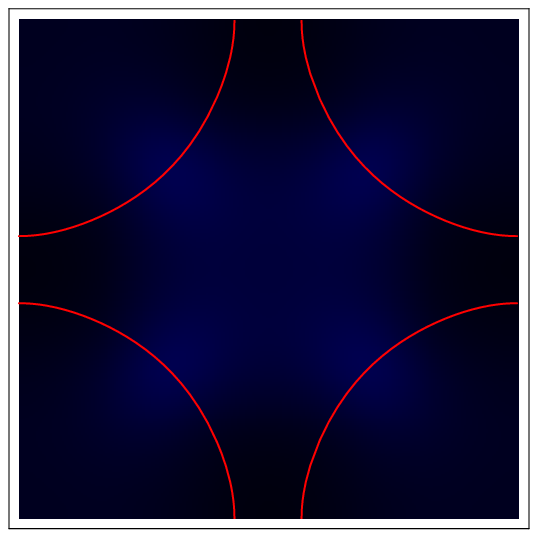}
\includegraphics[width=.45\columnwidth]{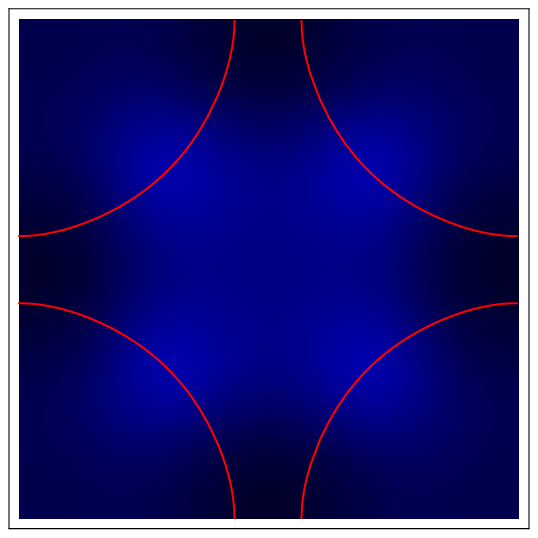}
\includegraphics[width=.45\columnwidth]{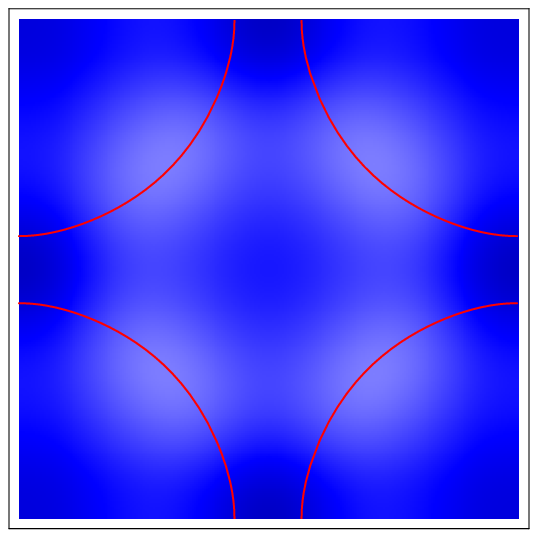}
\end{center}
\caption{(Color online) Same as Fig. \ref{fig:ElasticBZ02}, but
with less pronounced forward scattering, $\kappa =1.0$.
} \label{fig:ElasticBZ10}
\end{figure}

In Fig.~\ref{fig:ElasticBZ02} we show the imaginary part of the
$\tau_0$ and the $\tau_1$ contribution of the elastic self energy
to $\Gamma_{01} \equiv - {\rm Im} \, \Sigma_{\rm el,0} -
\Delta_{\bf k}/ \omega \; {\rm Im} \Sigma_{\rm el,1}$, that describes
the elastic quasiparticle scattering rate  (we have neglected the
$\Sigma_{el,3}$ term in (\ref{Eq:effective_scattrate}), which
vanishes on the Fermi surface, in order to highlight the behavior
there). In the case of a distinct forward scattering potential
with $\kappa = 0.2$, it is obvious that nodal quasiparticles can
be scattered at low energies while the scattering of antinodal
quasiparticles is strongly suppressed.
In Fig.~\ref{fig:ElasticBZ10}, we show the same quantity for a
more isotropic scattering potential with $\kappa=1$. Here the
elastic scattering rate for quasiparticles at the Fermi surface
shows only a weak anisotropy with just a slightly increased
scattering rate for nodal quasiparticles. With increasing
$\kappa$, approaching the isotropic limit, the elastic scattering
rate at the Fermi surface is fully determined by the imaginary
part of $\Sigma_{\rm el,0}$ while $\Sigma_{\rm el,1}$ vanishes.

\begin{figure}
\begin{center}
\includegraphics[width=.49\columnwidth]{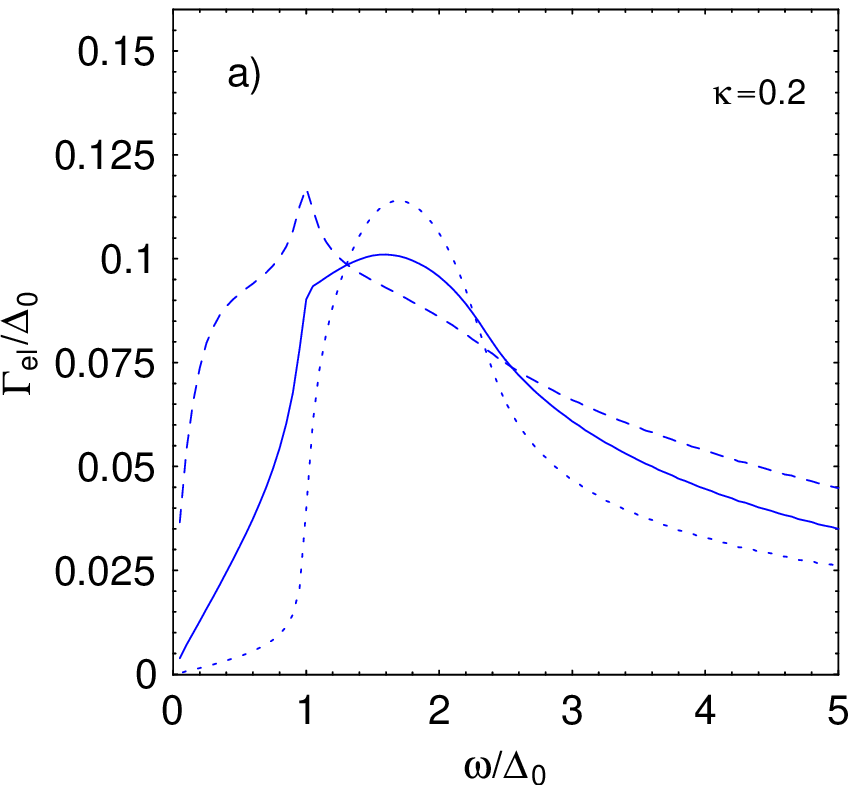}
\includegraphics[width=.49\columnwidth]{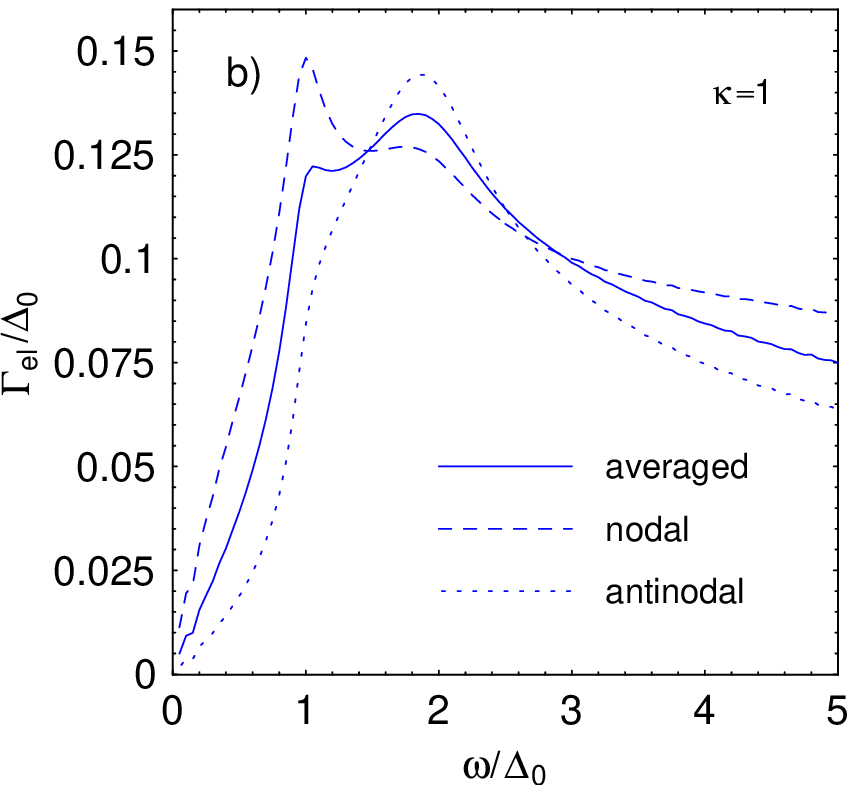}
\end{center}
\caption{(Color online) The elastic scattering rate  as a
 function of the quasiparticle energy $\omega$ for two different values of the inverse
scattering range $\kappa=0.2$ (a) and $\kappa=1$ (b). The solid
lines show the scattering rate averaged over the Fermi surface,
while the dashed and dotted lines show the scattering rate for
quasiparticles at the nodal and antinodal point of the Fermi
surface respectively.} \label{fig:elastic}
\end{figure}

In Fig.~\ref{fig:elastic}, we compare the energy dependence of the
nodal and antinodal elastic scattering rates for the two different
forward scattering parameters discussed above. We find for the
elastic scattering in the forward scattering limit ($\kappa =
0.2$) a high scattering rate for nodal quasiparticles already at
low energies compared to the gap $\Delta_0$, while the scattering
rate for quasiparticles at the antinodal points is strongly
suppressed and sets in only for higher
energies~\cite{TDahm:2005a}.  As previously discussed
\cite{ZHS04}, this behavior arises from a cancellation between the
``normal" $\Sigma_0$ and the ``anomalous" $\Sigma_1$ channels for
$\omega<\Delta_{\bf k}$ when the scattering is peaked in the forward
direction.  As expected, the scattering becomes more isotropic if
the range of the impurity potential is decreased ($\kappa = 1$)
and the frequency dependent scattering rates evaluated at
different points of the Fermi surface approach each other. In this
case, the scattering rate for nodal quasiparticles as well as the
Fermi surface averaged scattering rate shows a nearly linear
increase with frequency, reflecting the $d$-wave density of
states.

\subsection{Elastic off-diagonal scattering}
\label{subsec:tau1scatt}

 \vskip .2cm We mention briefly the notion
of off-diagonal or pairing disorder scattering by out of plane
impurities, claimed to be responsible in
Refs.~\cite{TSNunner:2005,JXZhu:2006,BMAndersen:2006,TSNunner:2006,ACFang:2006,MMaska:2007}
for the inhomogeneity in the gap magnitude measured by STS, as
well as for many of the correlations between various STS
observables. The scenario described in these works is that out of
plane dopant impurities, in addition to creating a screened
Coulomb potential, represent an atomic-scale modulation of the BCS
pairing interaction in the low-energy effective Hamiltonian, via a
modulation of local electronic structure parameters such as $t,t'$
or $J$, or of electron-phonon  interaction matrix elements. Since
the scattering by an induced off-diagonal potential
$\delta\Delta({\bf r})$ is necessarily weak on the scale of the Fermi
energy, it is again appropriate to use the Born approximation to
describe this scattering. The modulation of $\Delta({\bf r})$ observed
in experiment~\cite{SHPan:2001,CHowald:2001,KMLang:2002} is
however not pointlike, but has a characteristic wavelength of
25-30 \AA; within the picture of Ref.~\cite{TSNunner:2005}, this
arises because  the dopant atoms give rise to an atomic scale
modulation of the BCS pairing potential which causes larger,
coherence length size fluctuations in $\Delta_{\bf k}({\bf r})$.  This is
somewhat difficult to model as a quasiparticle scattering
potential in a disorder averaged calculation (see however
\cite{MHHettler:1999}), so we try in Appendix 1 to estimate the
crude effect of order parameter modulations on quasiparticle
scattering by allowing the order parameter to be modulated on the
4 bonds around the impurity, while assuming that it retains its
$d$-wave character\cite{AShnirman:1999}.  We find, not
surprisingly, that the scattering rate is largest at the antinode,
where it rises linearly in $\omega$ at zero temperature.

In an inhomogeneous system like BSCCO-2212, one might take $\delta
\Delta_{\bf k}$ to be of order $\Delta_0$ to reflect the large
distribution of observed gap values in the system.  The overall
scale of the scattering parameter $\Gamma^\Delta$ defined in the
Appendix would then vary as $\Delta_0^2$, leading to an averaged
local scattering rate of $\sim \Delta_0 \omega$.  It is tempting
to speculate that the measured strong correlation of the STS
scattering rate coefficient $\alpha({\bf r})$ defined in
(\ref{dwaveLDOS})-(\ref{proportion}) with the local $\Delta({\bf r})$
is due to this effect, but a much more rigorous treatment is
needed to establish this.

 The results in
Appendix 1  give a rough measure of the additional scattering rate
due to the $\tau_1$ or pairing disorder component. This would
apply, however, only in situations where the quasiparticles
explore many gap ``patches", such that a pairing disorder scenario
is valid.  However, as we argue below, this is the case only at
low energies, where the STS presents a picture of homogeneous
electronic structure.  At higher energies, such averaging is no
longer appropriate, and it would be interesting to consider in
more detail the effects of localization of quasiparticles within a
patch, see Refs.~\cite{TSNunner:2005} and \cite{ACFang:2006}.
Since a more microscopic approach is not available at the present
time, {\it and} a momentum average will clearly give an overall
linear-$\omega$ contribution to the elastic scattering rate
arising from this term (similar to the ordinary Coulomb
scattering), we will ignore its possible relevance for the time
being and return to it only in the discussion.
\subsection{Inelastic electron-electron scattering} \label{subsec:inelastic}

Besides the elastic scattering we will also take into account the inelastic
scattering that arises from the exchange of dynamic spin fluctuations. In the
random-phase approximation (RPA), the imaginary
part of the quasiparticle self-energy due to inelastic scattering
from the on-site Coulomb interaction $U$ can be written as
\cite{QSB94}
\begin{widetext}
\bea
-{\rm Im} \; \Sigma_{\rm inel}(\omega,{\bf k})
& = & \frac{3 U^2}{4}  \sum_q \left[ n(\omega-E_{\bf q}) + f(-E_{\bf q}) \right]
{\rm Im} \; \chi ({\bf k} - {\bf q},\omega-E_{\bf q}) \left( \tau_0 +
\frac{\epsilon_{\bf q}}{E_{\bf q}} \tau_3 + \frac{\Delta_{\bf q}}{E_{\bf q}} \tau_1 \right) \nonumber \\
 & + & \frac{3 U^2}{4} \sum_q \left[ n(\omega+E_{\bf q}) + f(E_{\bf q}) \right]
{\rm Im} \; \chi ({\bf k} - {\bf q},\omega+E_{\bf q}) \left(\tau_0 -
\frac{\epsilon_{\bf q}}{E_{\bf q}} \tau_3 - \frac{\Delta_{\bf q}}{E_{\bf q}} \tau_1
\right)\label{rpa_chi} \eea
\end{widetext}

\begin{figure}
\begin{center}
\includegraphics[width=.45\columnwidth]{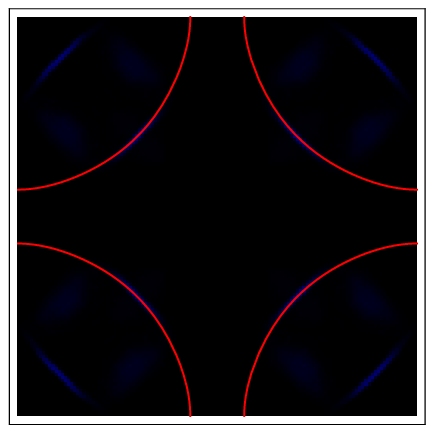}
\includegraphics[width=.45\columnwidth]{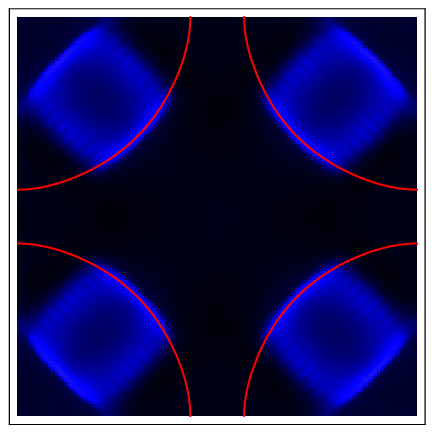}
\includegraphics[width=.45\columnwidth]{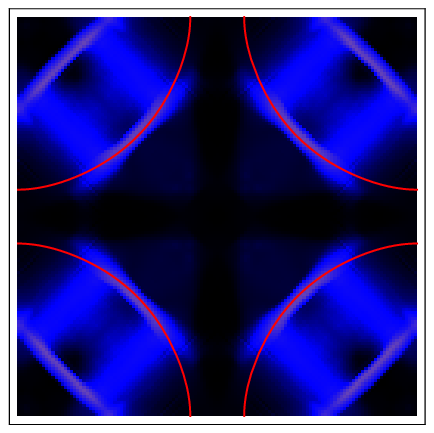}
\includegraphics[width=.45\columnwidth]{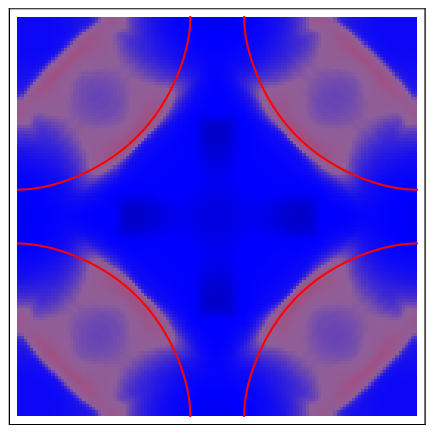}
\includegraphics[width=.45\columnwidth]{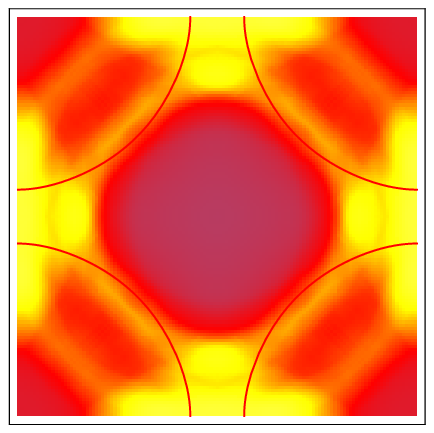}
\includegraphics[width=.45\columnwidth]{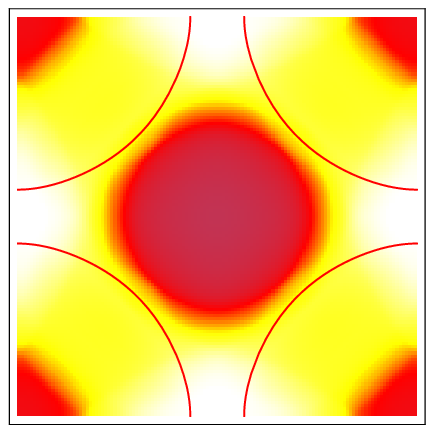}
\end{center}
\begin{flushleft}
\includegraphics[width=.45\columnwidth]{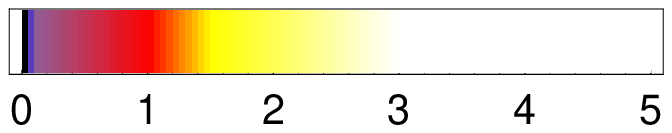}
\end{flushleft}
\caption{(Color online) The sum of the $\tau_0$ and the $\tau_1$
contributions to the inelastic scattering rate (which determine
the scattering near the Fermi surface) as a function of $k_x$ and
$k_y$ in the first Brillouin zone for different quasiparticle
energies $\omega$.  In the left column the   coupling constant is
taken to be $\bar U=2.20t$, which does not produce a well-defined
resonant mode.  In the right column we show results  for $\bar
U=2.36t$, which produces a resonant mode and an incommensurate
spin response. From top to bottom $\omega$ is chosen to be $\omega
= 0.5 \Delta_0$, $\omega = \Delta_0$ and $\omega =2.5 \Delta_0$.
The red line shows the Fermi surface. The color scale is the same
for all six plots and is given below. The units of the scattering
rate $\Gamma$ shown in the color bar are $\Delta_0$.}
\label{fig:InelasticBZ}
\end{figure}

In this expression $n(\omega)$ and $f(\omega)$ are the Bose and
Fermi distribution functions, $E_{\bf k} = \sqrt{\epsilon_{\bf k}^2 +
\Delta_{\bf k}^2}$ and the RPA expression for the spin susceptibility
$\chi$ is given by \be \chi ({\bf k},\omega) = \frac{\chi_0
({\bf k},\omega)}{1 - {\bar U} \chi_0 ({\bf k},\omega)}, \label{rpa_form}
\ee where $\chi_0$ is the bare BCS spin susceptibility

\bea
\lefteqn{ \chi_0 ({\bf q}, \omega) = } \nonumber \\
& & \sum_k \left\{ \frac{1}{2} \left[1 + \frac{\epsilon_{{\bf k}+{\bf q}}
\epsilon_{\bf k} + \Delta_{ {\bf k}+{\bf q}}\Delta_{\bf k}}{E_{{\bf k}+{\bf q}}E_{\bf k}} \right]
\frac{f(E_{{\bf k}+{\bf q}}) - f(E_{\bf k})}{\omega-(E_{{\bf k}+{\bf q}}-E_{\bf k})+ i0^+} \right. \nonumber \\
& & + \frac{1}{4} \left[1 -
\frac{\epsilon_{{\bf k}+{\bf q}}\epsilon_{\bf k}+\Delta_{{\bf k}+{\bf q}}\Delta_{\bf k}}{E_{{\bf k}+{\bf q}}E_{\bf k}}
\right] \frac{1 - f(E_{{\bf k}+{\bf q}})-f(E_{\bf k})}{\omega+(E_{{\bf k}+{\bf q}}+E_{\bf k}) + i0^+} \nonumber \\
& & \left. + \frac{1}{4}
\left[1-\frac{\epsilon_{{\bf k}+{\bf q}}\epsilon_{\bf k}+\Delta_{{\bf k}+{\bf q}}\Delta_{\bf k}}{E_{{\bf k}+{\bf q}}E_{\bf k}}
\right] \frac{f(E_{{\bf k}+{\bf q}})+f(E_{\bf k})-1}{\omega-(E_{{\bf k}+{\bf q}}+E_{\bf k}) +
i0^+ } \right\}. \nonumber \eea Note that in (\ref{rpa_chi})  the
coupling constant $U$ is in principle different from the effective
$\bar U$ that enters the denominator of the RPA expression
Eq.~(\ref{rpa_form}) for the spin susceptibility. Here $U$ and
$\tilde U$ will be treated as phenomenological parameters. Earlier
numerical calculations  have used $U={\bar U}=2.2 t$, to fit
microwave and thermal conductivity~\cite{DSH01,QHS96}. Here we
will set $U=2.2t$ and vary $\bar U$ in order to see what effects a
$\pi$-resonance produces.

\begin{figure}
\begin{center}
\includegraphics[width=.45\columnwidth]{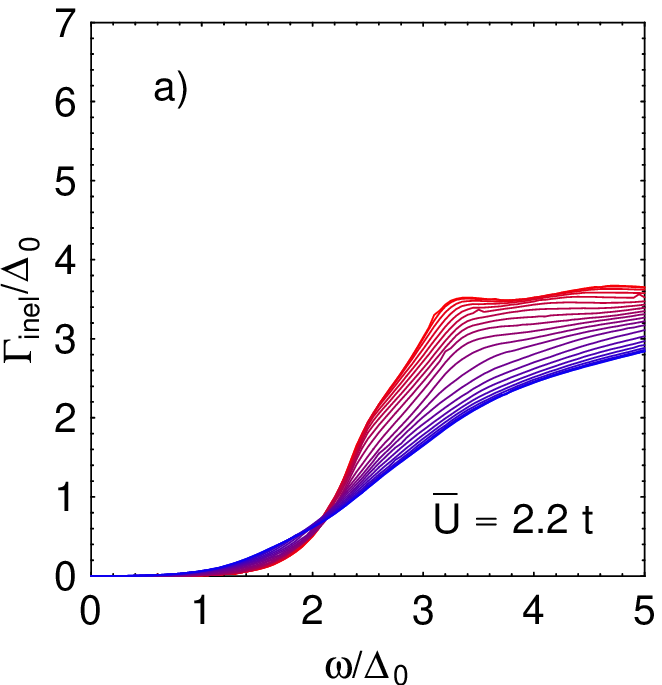}
\includegraphics[width=.45\columnwidth]{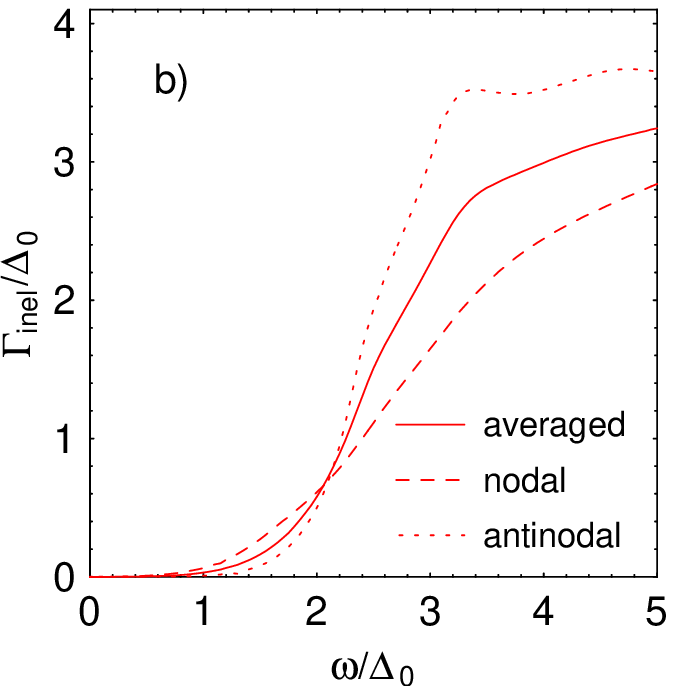}
\includegraphics[width=.45\columnwidth]{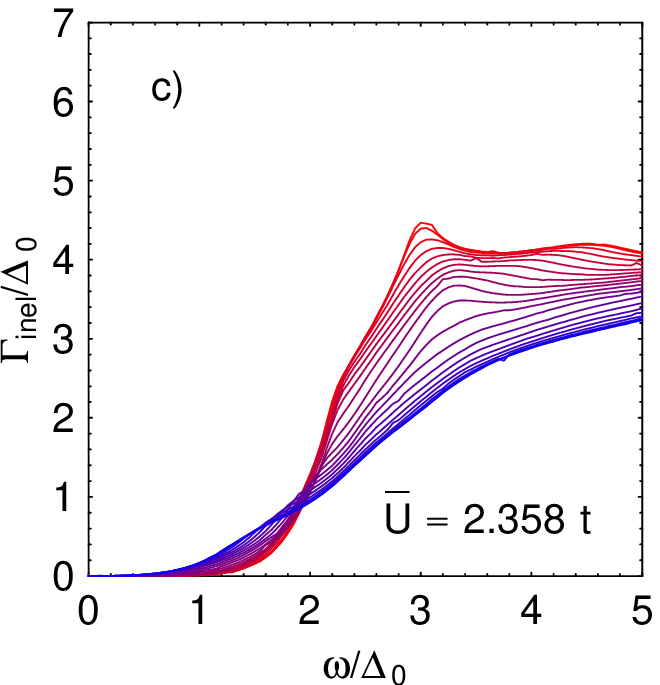}
\includegraphics[width=.45\columnwidth]{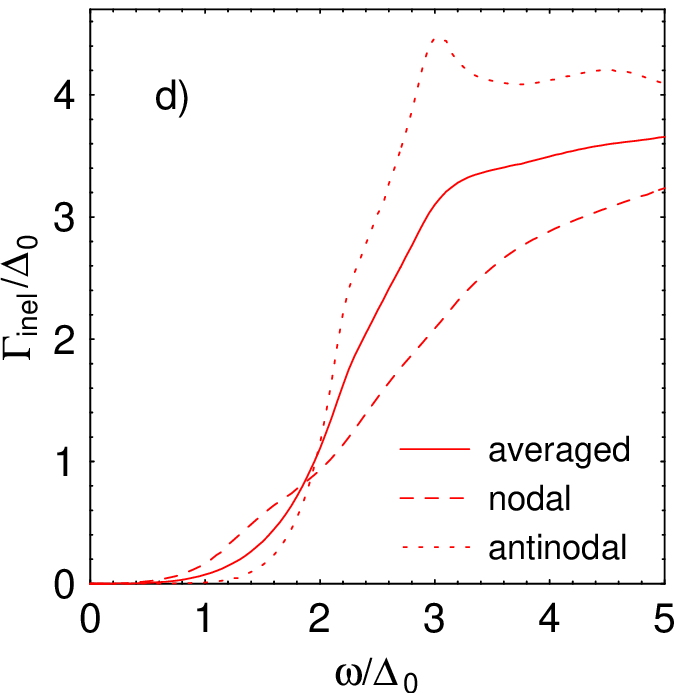}
\includegraphics[width=.45\columnwidth]{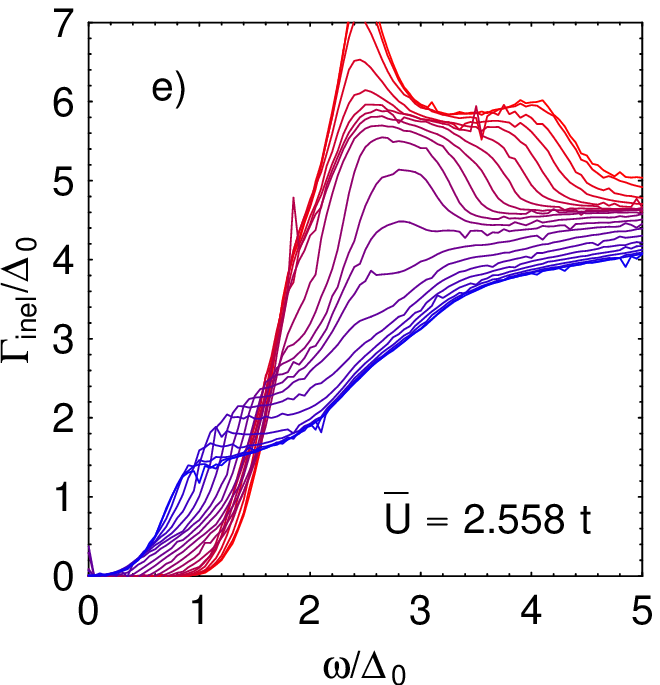}
\includegraphics[width=.45\columnwidth]{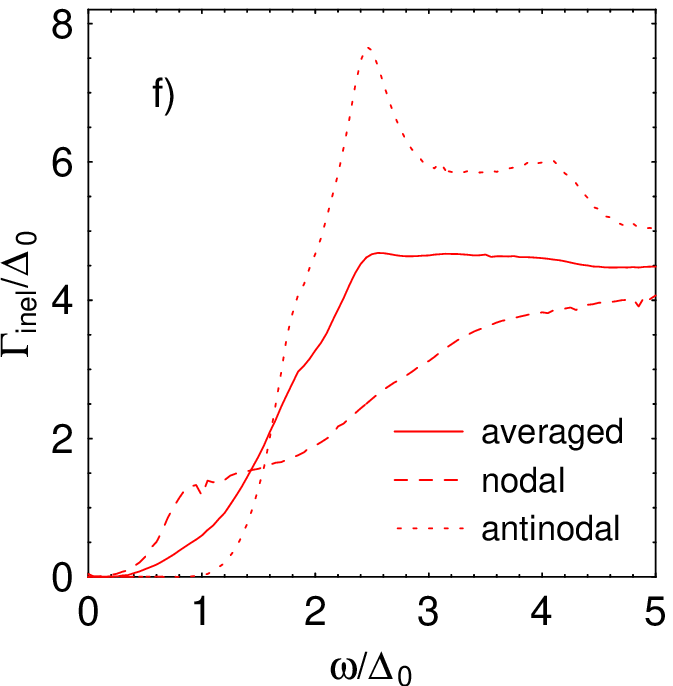}
\end{center}
\caption{(Color online) The inelastic scattering rate $\Gamma_{\rm
inel}({\bf k},\omega)$ for ${\bf k}$ on the Fermi surface as a function of
the quasiparticle energy $\omega$ for different values of
interaction parameter $\bar U=2.20t$ (a,b); 2.36$t$ (c,d); and
2.56$t$ (e,f).  Left panels: scattering rate at different ${\bf k}$
ranging in equal increments along Fermi surface from node (blue)
to antinode (red).  Right panels: The solid lines show the
scattering rate averaged over the Fermi surface, while the dashed
and dotted lines show the scattering rate for quasiparticles at
the nodal and antinodal point of the Fermi surface respectively. A
constant broadening of $0.1 \Delta_0$ has been added to regularize
the numerical calculation. }\label{fig:inel_scattrate}
\end{figure}
In Fig.~\ref{fig:InelasticBZ} the sum of the imaginary parts of
the $\tau_0$ and $\tau_1$ component of the inelastic scattering
rate, which determines the inelastic scattering of quasiparticles
at the Fermi surface, are shown in momentum space for the choice
$U=\bar U=2.2t$ (left column) and $\bar U=2.36t$ (right column).
Besides the strong scattering peaks at the Fermi surface, we find
equally strong signals near the corner of the Brillouin zone, that
are visible for all energies $\omega$ and reflect the enhanced
spin susceptibility at the nesting vectors connecting the two
opposite branches of the Fermi surface.

The energy dependence of the scattering rate at the  nodal and
antinodal points on the Fermi surface for $\bar U=2.2t$ is now
exhibited in Fig. \ref{fig:inel_scattrate}(a) and (b).  In
contrast to the elastic rate, the inelastic scattering rate at
$T=0$ is strongly suppressed for low energies and shows an
$\omega^3$ dependence for nodal
quasiparticles\cite{QSB94,QHS96,analytic,JPaaske:2000,DSH01} while
it vanishes exponentially below $\Delta_0$ for antinodal
quasiparticles.  (Note that in the reference clean noninteracting
quasiparticle system,  antinodal quasiparticles with
$\omega<\Delta_0$ do not exist; the typical or ``on-shell"
lifetime of an antinodal excitation is discussed below.) The
$\omega^3$ dependence of the inelastic scattering for nodal
quasiparticles can be understood as a combination of the
Fermi-liquid like $\omega^2$ dependence and the linear $\omega$
dependence of the $d$-wave density of states. At higher energies,
on the other hand, the broadening of the antinodal states in the
case of weaker interactions (Fig.~\ref{fig:inel_scattrate} (a) and
(b)) becomes somewhat larger than that of the nodal states, due to
the crossover to the normal state band structure.  In
Fig.~\ref{fig:inel_scattrate} (b) we have compared the energy
dependence at the node and antinode with the Fermi surface
average, which will be useful to us below. It is worthwhile noting
at this stage that the full Brillouin zone average of $\Gamma_{\rm
inel}({\bf k},\omega)$ (not shown) does not differ qualitatively from
the Fermi surface average.

Within the generalized RPA spin fluctuation approach to the
Hubbard model, it has been known for some time that if $\bar U$ is
adjusted to tune the system closer to the antiferromagnetic phase
transition, a resonant $S=1$ collective mode is pulled down below
the particle-hole continuum in the dynamical  spin susceptibility
$\chi''({\bf q},\omega)$ near ${\bf q}=(\pi,\pi)$. This mode has been
identified as the sharp spectral feature observed in neutron
scattering measurements at this
wavevector~\cite{earlycollectivemode}. The RPA  represents one of
several approaches which have been proposed to describe the
$\pi$-resonance which is observed in the cuprates
\cite{Eschrigreview}.  Specific applications to the neutron
response of  different cuprate materials have been given in Ref.
\cite{MRNorman:2007b}.

\begin{figure}
\begin{center}
\includegraphics[width=.43\columnwidth]{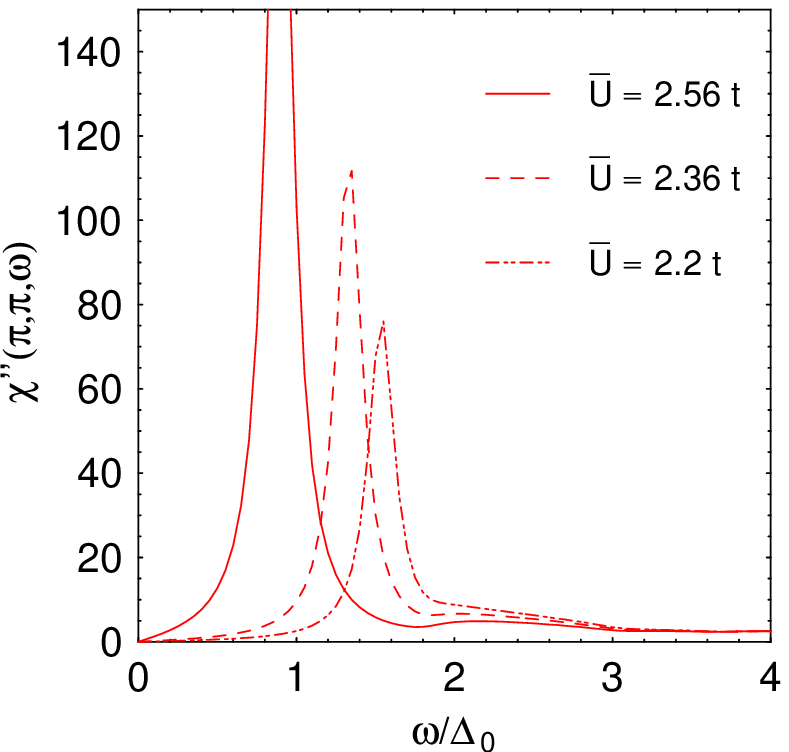}
\includegraphics[width=.43\columnwidth]{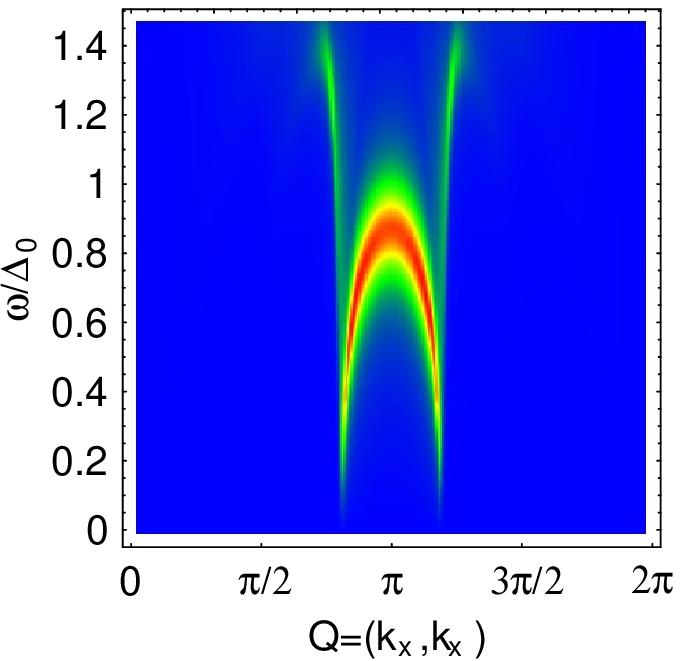}
\includegraphics[width=.11\columnwidth]{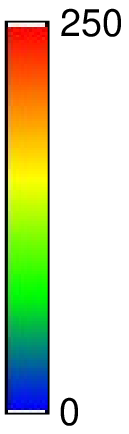}
\end{center}
\caption{(Color online) Left: imaginary part of dynamical
susceptibility $\chi''({\bf q},\omega) $ vs.\  $\omega$ for the three
values of $\bar U$ considered in the text.  Right: false color
plot of $\chi''({\bf q},\omega) $  for a cut through the Brillouin zone
along (110).
 } \label{fig:resonant_chi}
\end{figure}

Here we first discuss the basic kinematics of the collective mode
and consider the effect on the fermionic one-particle self-energy
within our model.  This has been considered already using
approximate models for the
susceptibility~\cite{MEschrig:2000,AAbanov:1999,AAbanov:2002,JFink:2006}
for purposes of comparing features of the neutron and ARPES
spectra. Within the RPA form Eq.~\ref{rpa_form}, a resonance in
the real part of the susceptibility may occur when \bea {\bar U} =
1/\chi_0({\bf Q},\omega_{res}). \eea  For our band parameters, and if
we require ${\bf Q}=(\pi,\pi)$, the resonant frequency occurs at
$\omega_{res}=0.9 \Delta_0$ for ${\bar U}=2.56$.   More generally,
for ${\bf Q}$ sufficiently close to $(\pi,\pi)$, it may be shown that
$\chi_0'' ({\bf Q},\omega_{res})$ vanishes at $T=0$.  In this case
there is a sharp collective mode contribution  of the form \bea
\chi_{res}''({\bf Q},\omega) = {\pi\over {\bar U}} \delta\left(1-{\bar
U} \chi_0'({\bf Q},\omega)\right) \eea to the imaginary part of the
spin susceptibility near ${\bf Q}$. In Figure~\ref{fig:resonant_chi},
we show how, with increasing interaction strength, the energy
position  of the collective feature in $\chi''$ shifts downward
and the intensity increases.  These aspects of the spin
fluctuation spectrum are then directly reflected in Fig.
\ref{fig:inel_scattrate}. Many of the features of the experimental
neutron spectrum on optimally doped YBCO\cite{neutron_expt} and
BSCCO\cite{BFauque:2007}, including the hourglass-like shape of
the neutron response in ${\bf q},\omega$ space (Fig.
\ref{fig:resonant_chi} (b)), as well a characteristic rotation of
the pattern of incommensurate ${\bf q}$-peaks as the energy is tuned
through the resonance are known to be captured by this type of
theory~\cite{IEremin:2005,Eschrigreview}.

To see how the collective mode affects the quasiparticle
scattering rate, we focus on the resonant part of $\chi''$. The
contribution to the effective interaction $V_{res}({\bf q},\omega)$ is
${3\over 2} U^2 \chi''_{res}({\bf q},\omega)$, giving rise to an
additional contribution to the total scattering rate

\bea \Gamma_{res}({\bf k},\omega)=-{1\over 2}\sum_{{\bf k}',s=\pm} \, {\rm Im}\, V_{res}({\bf k}-{\bf k}',\omega+s E_{{\bf k}'})  \\
\times\left( f(s E_{{\bf k}'})+n(\omega +s E_{{\bf k}'})\right) \left[1 - s
{\epsilon_{\bf k}\epsilon_{{\bf k}'} + \Delta_{\bf k}\Delta_{{\bf k}'}\over \omega
E_{{\bf k}'}} \right].\nonumber\eea

At the node ${\bf k}={\bf k}_N$, the expressions simplify considerably and
we find at $T=0$ that

\bea \Gamma_{res}({\bf k}_N,\omega) &\simeq& {3\pi U^2\over 4 {\bar U}}
\sum_{{\bf k}'}
   \delta\left(1-{\bar U} \chi_0'({\bf k}-{\bf k}',\omega
- E_{{\bf k}'})\right)\nonumber\\&& \times \theta(\omega-E_{{\bf k}'}). \eea
The integrand contributes provided $\omega = \omega_{res} +
\sqrt{\epsilon_{{\bf k}'}^2+\Delta_{{\bf k}'}^2}$, with ${\bf k}'={\bf k} - {\bf Q}$. Since
for our band parameters ${\bf k}'={\bf k}_N -(\pi,\pi)$ is far from the
Fermi surface, the contribution to $\Gamma$ from the magnetic
resonance mode will arise at frequencies $\omega=\omega_{res} +
|\epsilon_{{\bf k}'}|$   larger than $\omega_{res}$ itself by an amount
which is of order a fraction of the Fermi energy, much larger than
$\Delta_0$.

For the antinodal points, or rather  the nearby ``hot spots"
${\bf k}_{HS}$ where the Fermi surface crosses the antiferromagnetic
zone boundary, the wave vector ${\bf k}' = {\bf k}_{HS} - (\pi,\pi)$ is also
on the Fermi surface.  However the off-diagonal parts of the
coherence factors must now also be included,  so the resonant
contribution becomes \bea \Gamma_{res}({\bf k}_N,\omega) &\simeq &
{3\pi U^2\over 4 \bar U} \sum_{{\bf k}'}
   \delta\left(1-\bar U \chi_0'({\bf k}-{\bf k}',\omega
- E_{{\bf k}'})\right)\nonumber\\&& \times
\theta(\omega-E_{{\bf k}'})\left[1-{\Delta_{HS}\over \omega}\right],
\eea where $\Delta_{HS}\approx \Delta_0$ is the value of the order
parameter at the hot spot ${\bf k}_{HS}$.  Thus there is a contribution
to the resonant part of the scattering rate at the hot spot when
$\omega\simeq \omega_{res} + \Delta_0$.  This additional
scattering has been claimed to be responsible for part of the
``peak-dip-hump" structure seen in ARPES near the antinode at low
temperatures in the superconducting
state~\cite{MEschrig:2000,AAbanov:1999}. However, since
$\omega_{res}$  is also of order $\Delta_0$, it does not appear as
though the resonant electron-electron scattering
 can by itself play a significant role in the electronic scattering rate for energies at or
below the coherence peak energy $\Delta_0$.

Nevertheless, the enhanced inelastic scattering will play a role
in measured STS properties.  The tuning of the system closer to
the antiferromagnetic instability has the effect of enhancing the
overall scattering rate, such that the states near the nodes now
decay more rapidly.  In Figure~\ref{fig:inel_scattrate}(c)-(f), we
show the evolution of the energy-dependent scattering rate for
states along the Fermi surface for slightly larger $\bar U$. 
The contribution of the collective mode is easiest to discern in
(e)-(f), for the case $\bar U=2.56$.  The scattering in the midgap
range near the node has risen considerably, although
near the antinode it is still exponentially suppressed at low
energies.  The collective  mode results in a contribution which
rises near $\omega\simeq \omega_{res}+\Delta_0\approx 1.9
\Delta_0$ at the antinode, as expected from the above discussion,
which  then peaks at about $2.5\Delta_0$.  There is no obvious
resonant contribution at higher energies in the nodal direction,
but its existence can be deduced by considering the evolution  of
the antinodal peak as one moves away from the antinode: as seen in
the figure, it moves to higher energies as expected and  broadens
due to the additional decay channels. Above an energy of order
1.5$\Delta_0$, the antinodal scattering rate is seen to become
much larger than the nodal scattering rate. 
This enhancement is directly related to the opening of the
superconducting gap and the creation of the resonant mode.

The intermediate case $\bar U=2.36$
(Fig.~\ref{fig:inel_scattrate}(c)-(d)) also shows clear evidence
of  enhanced interactions, and vestiges of the mode features. The
effect on $\Gamma_{\rm inel}$ at low energies is primarily
quantitative: the inelastic scattering near the node becomes
significant in the midgap range compared to the weaker case
(a)-(b), and is already  a fraction of $\Delta_0$ near the node at
the gap frequency $\omega=\Delta_0$.  This will play a role in our
discussion of STS experiments, because the extracted quasiparticle
scattering rate is of this order for optimally doped samples.
There is further independent  evidence in the work of Fauqu\'e et
al. \cite{BFauque:2007} to suggest that the intermediate
interaction strength case we considered here is most relevant for
experiments on optimally doped BSCCO. This is because in Ref.
\cite{BFauque:2007}, the collective mode energy is measured to be
42 meV, while the average gap is extracted as 35 meV. This ratio
of about 1.2 is achieved in our model for interaction strength
$\bar U=2.36$; changing this parameter moves the resonance up or
down relative to the gap edge, as seen in Fig.
\ref{fig:resonant_chi}(a).

\subsection{Total scattering rate} \label{subsec:total}
\label{subsec:total}
\begin{figure}
\begin{center}
\includegraphics[width=.47\columnwidth]{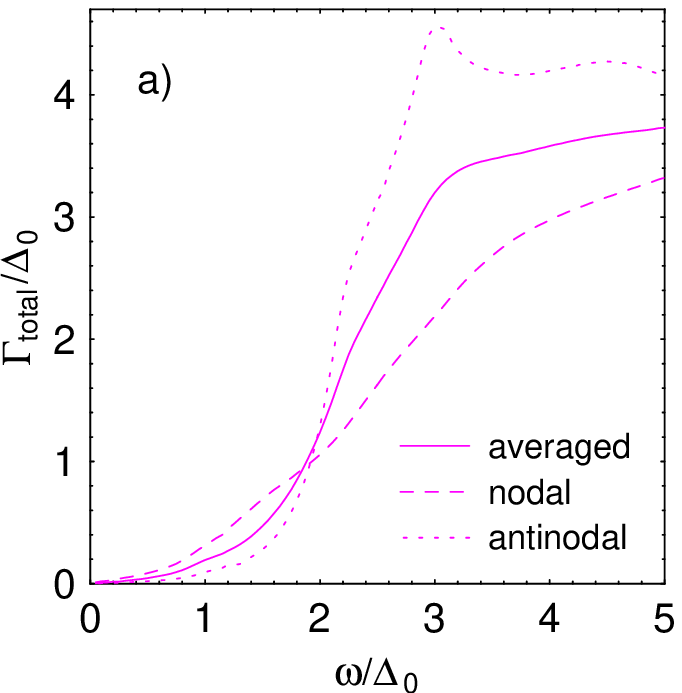}
\includegraphics[width=.51\columnwidth]{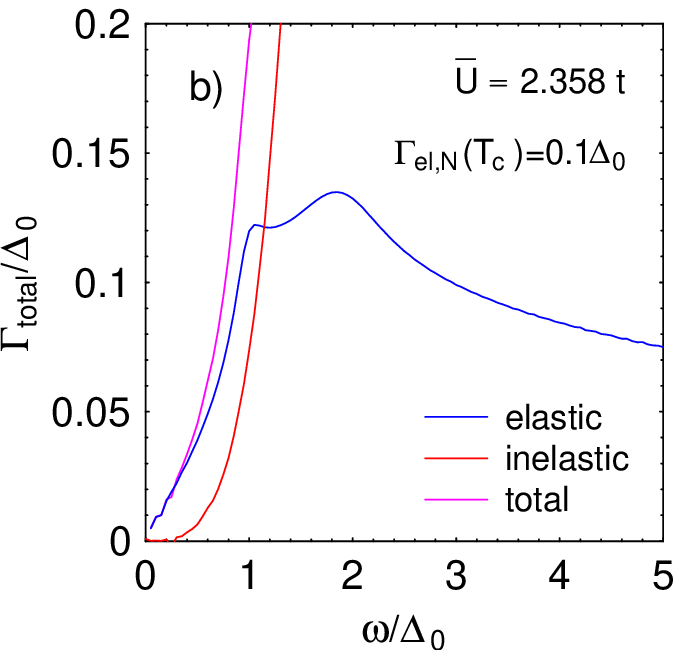}
\includegraphics[width=.47\columnwidth]{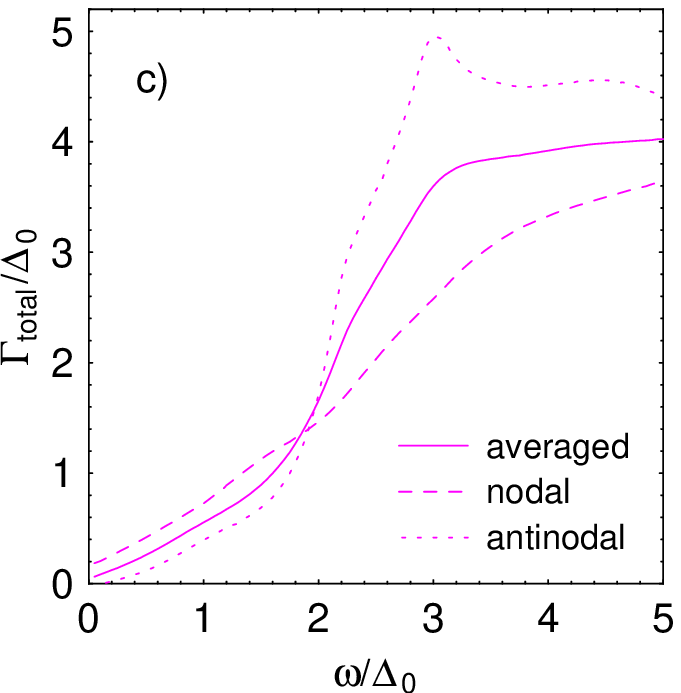}
\includegraphics[width=.51\columnwidth]{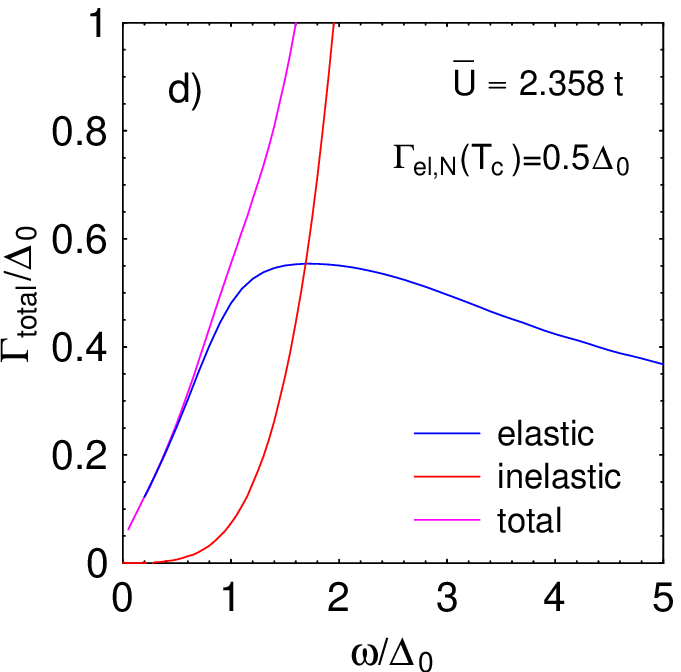}
\includegraphics[width=.47\columnwidth]{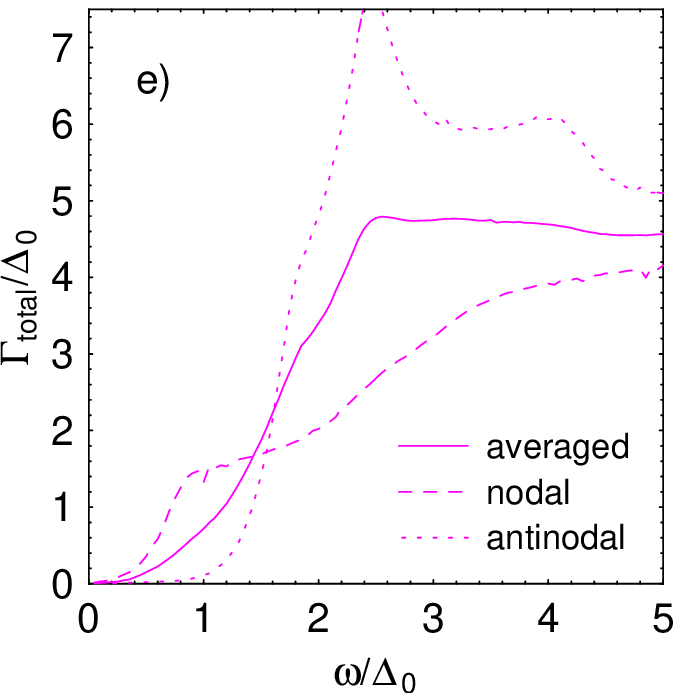}
\includegraphics[width=.51\columnwidth]{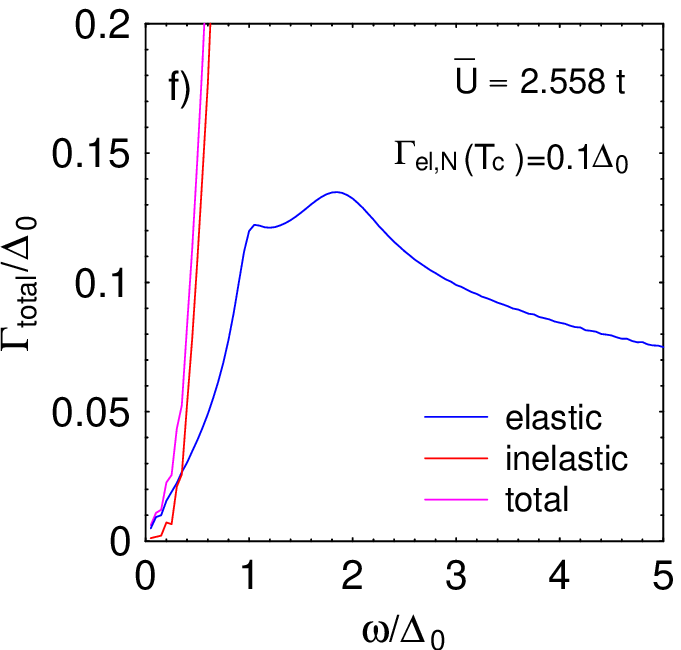}
\end{center}
\caption{(Color online) The elastic, inelastic, and total
scattering rates as a function of the quasiparticle energy
$\omega$ for several assumptions about the strength of each
scattering channel.  In the left panels, the total rate
$\Gamma_{tot}$ at the antinode ${\bf k}_A$ (node ${\bf k}_N$) is plotted as
a dotted  (dashed) line.  The average over the Fermi surface of
$\Gamma_{tot}({\bf k},\omega)$ is plotted as the solid line. In the
right panels, the contribution of the elastic (blue) and inelastic
(orange) to the total (magenta) Fermi surface averaged rates is
exhibited separately.  (a)-(b) are for the case $\bar U=2.36t$
with $\Gamma_{el,N}(T_c)=0.1\Delta_0$. (c)-(d) Same as (a)-(b) but
with $\Gamma_{el,N}(T_c)=0.5\Delta_0$. (e)-(f) Same as (a)-(b) but
with $\bar U=2.56t$.  } \label{fig:combined}
\end{figure}

To include both elastic and inelastic  scattering effects, we
neglect interference processes between electron-electron
collisions and impurity scattering entirely, and approximate the
total scattering rate by \be \Gamma_{\rm tot}=\Gamma_{\rm
el}+\Gamma_{\rm inel}. \label{total}\ee In
Fig.~\ref{fig:combined}, the total scattering rate for
quasiparticles at the nodal point of the Fermi surface (dashed) is
compared to the scattering rate for quasiparticles at the
antinodal point of the Fermi surface (dotted) as well as the
average scattering rate including only the quasiparticles at the
Fermi surface, \bea \Gamma_{avg}(\omega)=\langle
\Gamma_{tot}({\bf k},\omega)\rangle_{FS}, \label{Eq:avgGamma}\eea where
$\langle\rangle_{FS}$ is a Fermi surface average.  The total
scattering rate is seen in Fig.~\ref{fig:combined} (a)-(b) to be
dominated for $\omega \lesssim 2\Delta_0$ by the linear energy
dependence of the elastic part, while at higher energies
$\omega>3\Delta_0$ it reflects the quasilinear energy dependence
of the inelastic scattering rate in the classical regime. The
slope for low energies and the slope for high energies are not
related to each other and depend on the individual parameters used
in the two models.

For completeness, in Fig.~\ref{fig:combined} (b) the total
scattering rate averaged over the Fermi surface is shown together
with its two averaged contributions from  elastic and inelastic
scattering processes. It shows qualitatively the same energy
dependence as the Brillouin zone averaged rate (not shown) but has
a somewhat larger overall magnitude. This is due to the fact that
the Brillouin zone average includes regions of the  zone away from
the Fermi surface with small scattering rates.

\section{Local Green's function}
\label{sec:localG}

In this section we calculate the exact local Green's function
including impurity and spin-fluctuation scattering, and exhibit
the LDOS for varying amounts of disorder and different interaction
strengths. Our first goal is to see if the use of a local
self-energy, which we define in an ad hoc way, can provide a good
description of the exact LDOS calculated with the full
self-energies.   Secondly, we compare these approximations to
 the linear-$\omega$ ansatz, given
by Eqs.~\ref{dwaveLDOS}-\ref{proportion}, employed by the Cornell
group to fit their data~\cite{JAlldredge:2007}.  We would like to
see if an approximate local self-energy may be constructed from
the local Green's function, and to what extent the true
self-energy yields an LDOS similar to that obtained from the
linear-$\omega$ scattering rate.

 The local Green's function $\underline G({\bf r},{\bf r},\omega)$  in the
presence of elastic and inelastic scattering is given, within a
region of size of order $\ell$, as the Fourier transform of the
momentum dependent Green's function $\underline G({\bf k},\omega)$ \be
\underline G({\bf r},{\bf r},\omega)=\sum_k \underline G({\bf k},\omega)
\label{Eq:localG} \ee Here the full momentum dependent Green's
function is calculated from \be \underline G({\bf k},\omega) =
\frac{\tilde{\omega} \tau_0 + \tilde{\epsilon}_{\bf k} \tau_3 +
\tilde{\Delta}_{\bf k} \tau_1} {\tilde{\omega}^2-\tilde{\epsilon}_{\bf k}^2
- \tilde{\Delta}_{\bf k}^2} \ee and $\tilde{\omega}$,
$\tilde{\epsilon}_{\bf k}$ and $\tilde{\Delta}_{\bf k}$ are the renormalized
quasiparticle energy, band structure and gap in the presence of
the elastic and inelastic self-energy calculated in the previous
section. The LDOS is then simply given by \bea
N({\bf r},\omega)=-{1\over 2\pi} {\rm Tr}\, \left[(\tau_0+\tau_3)
\underline G ({\bf r},{\bf r};\omega)\right].\eea

\begin{figure}
\begin{center}
\includegraphics[width=.49\columnwidth]{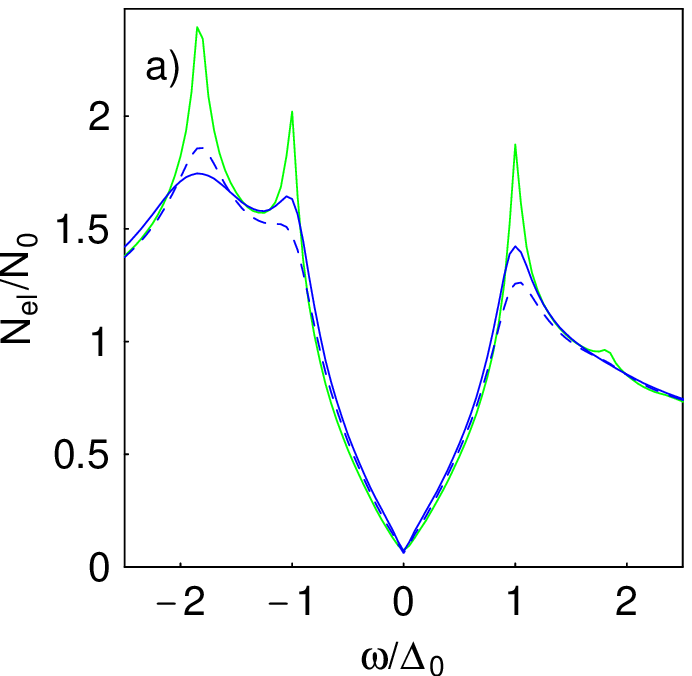}
\includegraphics[width=.49\columnwidth]{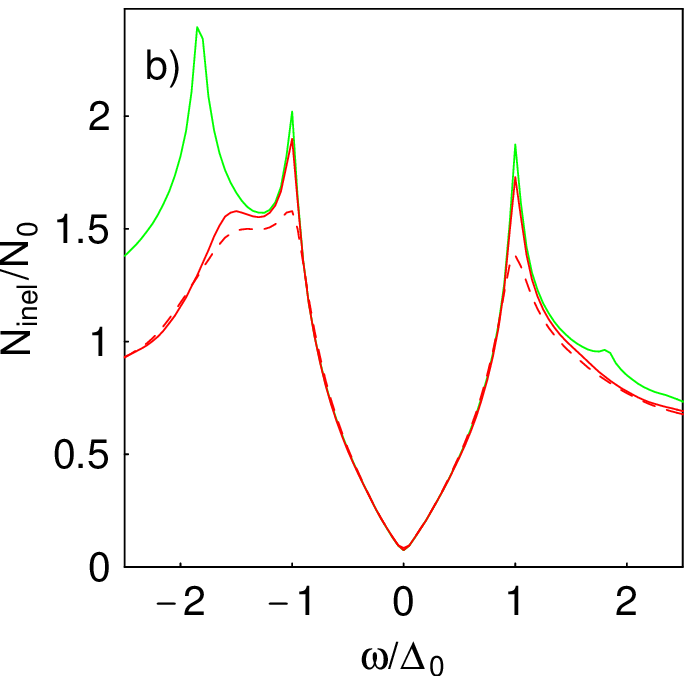}
\includegraphics[width=.49\columnwidth]{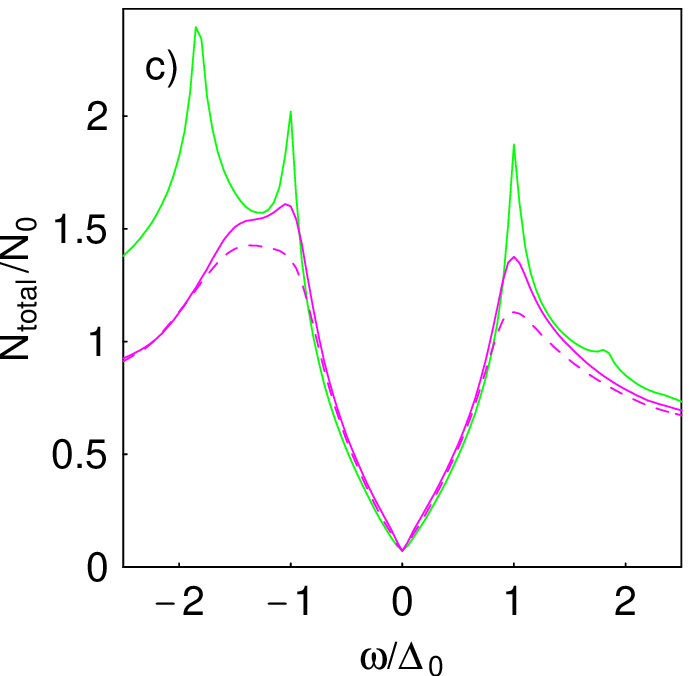}
\includegraphics[width=.49\columnwidth]{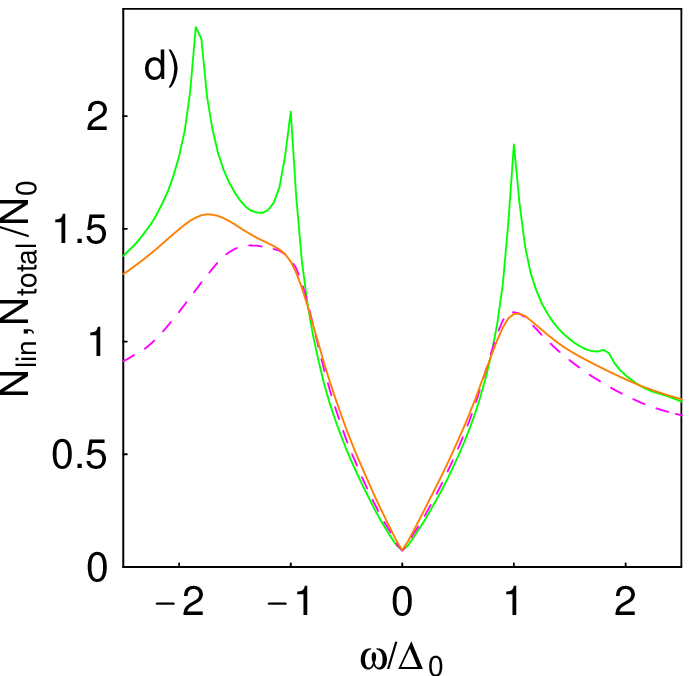}
\includegraphics[width=.49\columnwidth]{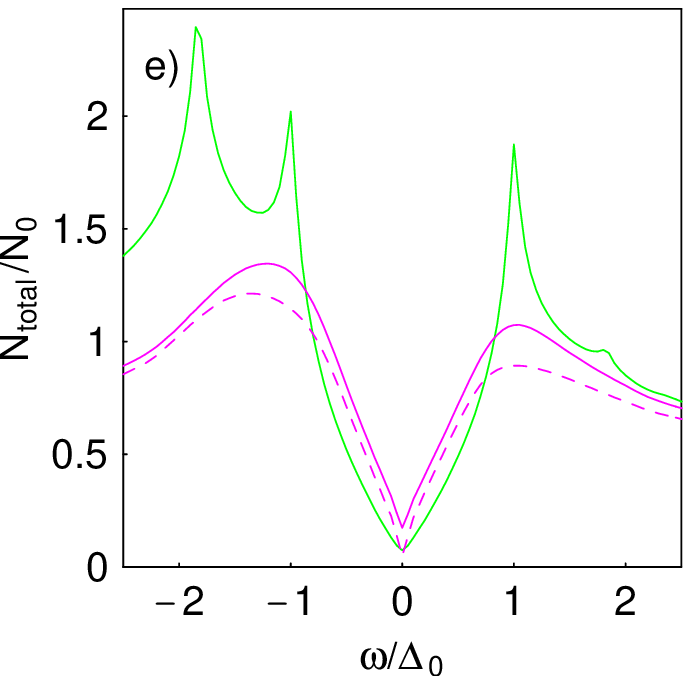}
\includegraphics[width=.49\columnwidth]{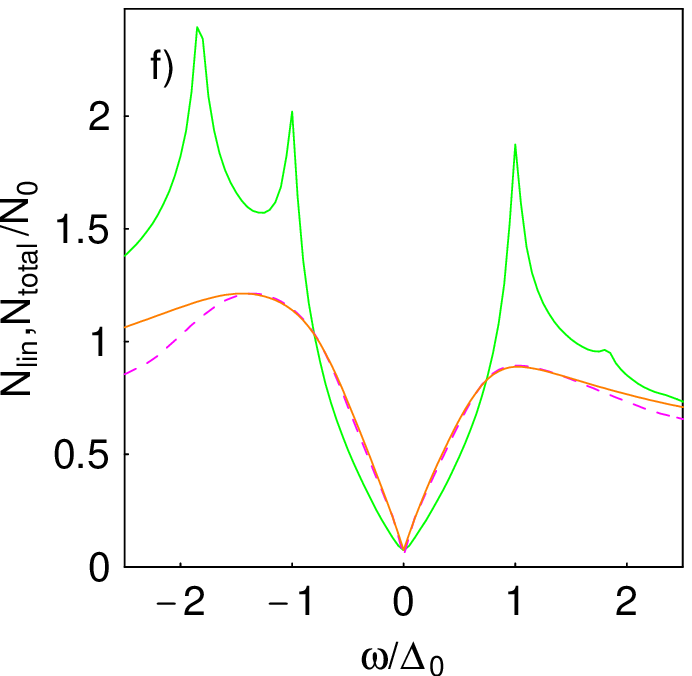}
\end{center}
\caption{(Color online) The local density of states (LDOS) $N_{\rm
total}(\omega)$ normalized to the Fermi level density of states
$N_0$ for a $d$-wave superconductor with scattering processes of
various types included. In all panels, green line shows LDOS for
pure system, blue (a) gives the LDOS for pure elastic scattering
with $\Gamma_{\rm el}(T_c)=0.1\Delta_0$,   red curve (b) gives
pure inelastic scattering with $U=2.36$, and magenta  (c) includes
both elastic and inelastic processes. Solid lines are calculated
with the full momentum dependent self-energy, while the dashed
curves used an effective momentum averaged scattering rate defined
in (\ref{Eq:avgGamma}). In (d) the LDOS  in the presence of
impurity and spin fluctuation scattering, with parameters from (a)
and (b), is compared to the best  fit $N_{lin}(\omega)$ (orange)
using a linear scattering rate  of the form given in
Eqs.~(\ref{dwaveLDOS}-\ref{proportion}) as extracted from STS data
in Ref. \onlinecite{JAlldredge:2007}. Figs. (e) and (f) show the
LDOS for a dirtier system with $\Gamma_{\rm el}(T_c)=0.5\Delta_0$
(line types same as in (a)-(d)).
 }\label{fig:LDOS}
\end{figure}

The results of these calculations including the full self-energies
are shown as the solid lines in Fig.~\ref{fig:LDOS} for (a) pure
elastic scattering, (b) pure inelastic scattering  and (c) for the
total scattering rate. The green curve shows for comparison the
the LDOS for a clean $d$-wave superconductor. The high slope of
the elastic scattering rate leads to a linear energy dependence of
the total rate up to a value of $\Delta_0$ and a strong
suppression of the coherence peaks. The inelastic scattering rate,
which is strongly suppressed for low energies, begins to
significantly affect the LDOS only at energies $\omega \gtrsim
\Delta_0$ for the slightly off-resonant case $\bar U=2.36t$ case
shown, and has qualitatively no influence on the low energy
spectrum. In the same figure panels, we compare these results to
calculations with the ad hoc ``local" scattering rate
$\Gamma_{avg}(\omega)$ 
which we simply add as an imaginary part to the quasiparticle
energy, \be \underline G[{\bf r},{\bf r},\omega + i \Gamma_{avg}(\omega)]
= \sum_k \underline G[{\bf k},\omega+ i \Gamma_{avg}(\omega)].\ee As
can been seen, this approximation (dashed lines) leads to a
slightly lower spectral weight for low energies but exhibits also
a quasilinear energy dependence for the local density of states up
to an energy of $\Delta_0$. Finally we fit the effective momentum
averaged scattering rate with the linear energy dependence of
Eq.~\ref{proportion} and we find for low energies excellent
agreement with the approximation of a momentum averaged scattering
rate (orange).

Thus we conclude that for energies $\omega\ltwid\Delta_0$, there
is a quasilinear increase of $\Gamma$ arising from elastic
scattering processes and in calculating the LDOS it can be modeled
by the linear form Eq.~(\ref{proportion}). On the other hand, the
STS local scattering rate is also enhanced by inelastic scattering
processes when $\omega\gtwid\Delta_0$.  Despite the fact that this
destroys even the approximate linearity of the total scattering
rate (see Fig. \ref{fig:combined}), the momentum average inherent
in the local measurement implies that it is very difficult to
distinguish the true functional form of $\Gamma$ in the impurity
plus spin fluctuation model from the phenomenological linear
ansatz of Ref. \cite{JAlldredge:2007}.  We return to this point in
Section~\ref{sec:ARPES}  below.

The  results in Fig. \ref{fig:LDOS} are  qualitatively similar to
 earlier attempts to fit STS data by assuming a model self-energy
dressing a BCS-like $d$-wave Green's function. In particular,
Hoogenboom et al.~\cite{BWHoogenboom:2003} and de Castro et
al.~\cite{GLdeCastro:2007} compared data on BSCCO to self-energy
models with constant scattering rates, a marginal Fermi liquid
model~\cite{CMVarma:1989}, and  the model of
Ref.~\cite{MEschrig:2000} describing a coupling to a
phenomenological collective spin mode.  They concluded that
details of the peak-dip hump structure observed in some spectra
required the collective mode, and identified features at sum and
difference energies of the gap and collective mode frequency. Here
 we have attempted no detailed fits of STS data, as we
are primarily interested in  exploring the general notion of a
local scattering rate.  We note that Fig.~\ref{fig:LDOS} shows, as
in Refs.~\cite{BWHoogenboom:2003,GLdeCastro:2007}, the broadening
of the van Hove singularity by inelastic scattering, and that
out-of-plane elastic scattering may play a role as well.
Peak-dip-hump structures similar to experiment are seen in some
cases in Fig.~\ref{fig:LDOS}, but more investigation is needed to
determine whether these aspects of the experimental spectra
correspond to scattering effects of the type considered here.

\section{Comparison with ARPES spectral function}
\label{sec:ARPES}

\begin{figure}
\begin{center}
\includegraphics[width=.49\columnwidth]{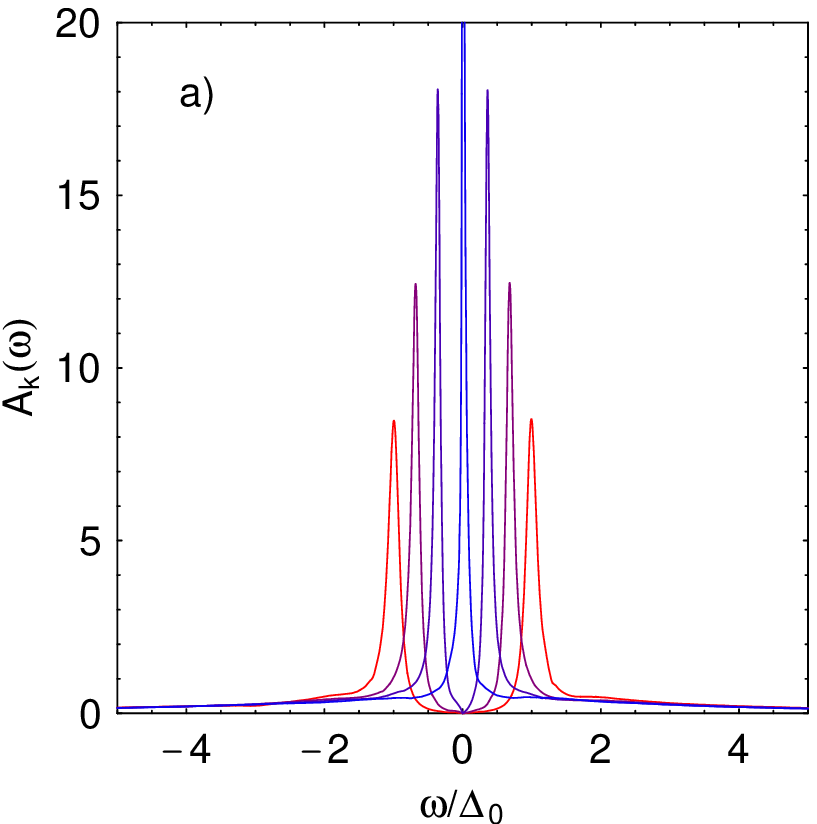}
\includegraphics[width=.49\columnwidth]{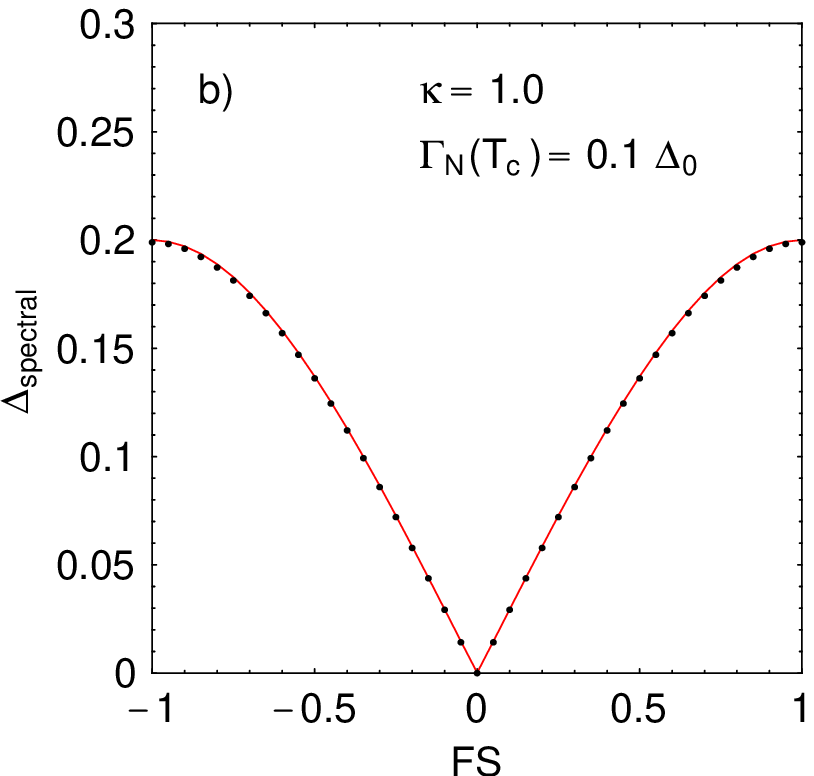}
\includegraphics[width=.49\columnwidth]{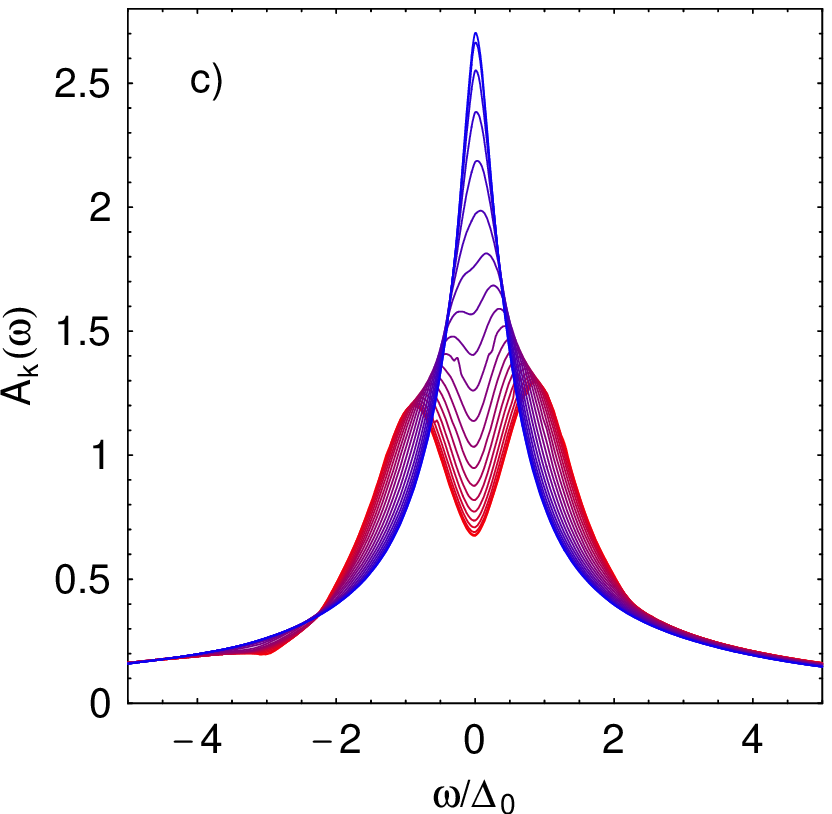}
\includegraphics[width=.49\columnwidth]{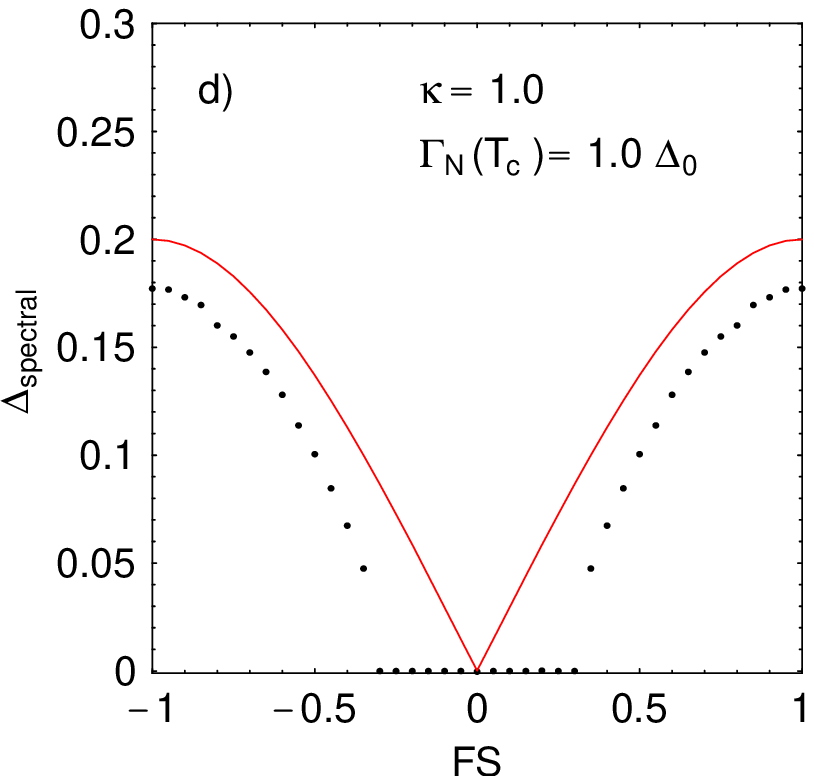}
\end{center}
\caption{(Color online) Left: spectral function $A({\bf k},\omega)$ for
${\bf k}$ on the Fermi surface at $T=0$ with elastic scattering
parameters $\kappa=1$ and  (a) $\Gamma_{el,N}(T_c)=0.1\Delta_0$
and (c) $\Gamma_{el,N}(T_c)=\Delta_0$. and with the inelastic
scattering rate calculated with $\bar U = 2.36t$.  Color ranges in
equal increments from blue (node) to red (antinode).  Only four
${\bf k}$ points are shown in (a) due to narrowness of peaks. Right:
spectral gap as determined by one-half the energy difference
between the two maxima of $A({\bf k},\omega)$ for (b)
$\Gamma_{el,N}(T_c)=0.1\Delta_0$ and (d) $=\Delta_0$,
respectively. Red curve in both figures indicates pure $d$-wave
gap in the absence of scattering.}\label{fig:spectral}
\end{figure}

The model of the quasiparticle scattering rate put forward above
has implications for other quantities besides the STS conductance,
and we may learn something from the comparison.  In particular, as
mentioned in the introduction, ARPES has reported superconducting
state relaxation rates which are  many times larger than those
measured by STS.   A naive application of these ARPES rates
 in the superconducting state would lead
to even broader  LDOS spectra than those in Fig.
\ref{fig:LDOS}(e)-(f), in contradiction to experiment.  The ARPES
signal is given in the sudden approximation by \bea I({\bf k},\omega) =
|M|^2 A({\bf k},\omega)f(\omega), \eea where $M$ is an ARPES matrix
element, which we take to be constant, $A({\bf k},\omega)=-(1/2\pi){\rm
Im\, Tr}\,(\tau_0+\tau_3)\underline G({\bf k},\omega)$, and $f$ is the
Fermi function. Here we calculate the full electron spectral
function $A({\bf k},\omega) $ within our simple model, for different
points on the Fermi surface and for different scattering rates,
with the aim of describing some qualitative phenomena and
comparing with STS. We note that the Fermi surface itself is, in
the presence of interactions and scattering, not identical with
the noninteracting pure one, but rather is the solution in ${\bf k}$
space of the equation \bea {\rm Re}\,\tilde\epsilon_{\bf k} =
\epsilon_{\bf k}  +{\rm Re}\, (\Sigma_3^{el}+\Sigma_3^{inel})=0. \eea
Here the chemical potential is determined to maintain a given
filling. We have verified that this does not affect the broadening
and in the calculation of the spectral functions shown in
Fig.~\ref{fig:spectral} the renormalization of the Fermi surface
by Re $\Sigma$ was neglected.

~

In Fig.~\ref{fig:spectral} (a),(c), we show the evolution of the
low-temperature spectral function along the Fermi surface,
beginning with the (blue) peak centered at $\omega=0$ as expected,
and ending with two clear (red) peaks at the antinode.  Close to
the node, some spectra still have their maxima at the Fermi
surface due to scattering effects.  In the pseudogap state of the
cuprates, a similar phenomenon is observed:  the maximum of the
spectral function remains at the Fermi surface over some range of
${\bf k}$ values centered on the  nodal point; the pseudogap itself is
visible as a double-peak structure only some distance away from
the nodal point.  The range of ${\bf k}$ where the single maximum is at
the Fermi energy is called the Fermi ``arc", and it evolves
continuously into the full Fermi surface in the normal state as
the pseudogap disappears.  It has been pointed out by several
authors~\cite{JGStorey:2007,AVChubukov:2007,MRNorman:2007} that
the arc phenomenon can be explained trivially by appealing to an
energy and/or temperature-dependent scattering rate, which for
sufficiently large scattering broadens the two-peak spectral
function one would expect in the presence of a spectral gap into a
single peak centered at the Fermi level.  A similar phenomenon
occurs here in the superconducting state, as shown in
Fig.~\ref{fig:spectral} (b),(d), where the effective gap,
determined from the position of the spectral function peak, is
plotted along the Fermi surface, exhibiting a finite range of ${\bf k}$
points where it vanishes.  For a constant scattering rate
$\Gamma$, the criterion determining the angular position of the
end of the arc is $\Gamma = a\Delta_{\bf k}$, where $a$ is a constant
of order unity, equal to $\sqrt{3}$ for a $d$-wave superconductor
with circular Fermi surface and $\cos 2\phi$ order
parameter\cite{MRNorman:2007}.  Note the effective gap determined
by ARPES in this manner does not correspond to the renormalized
order parameter in the theory, as pointed out by Sensarma et
al.~\cite{RSensarma:2007}.

\begin{figure}[t]
\begin{center}
\includegraphics[width=.95\columnwidth]{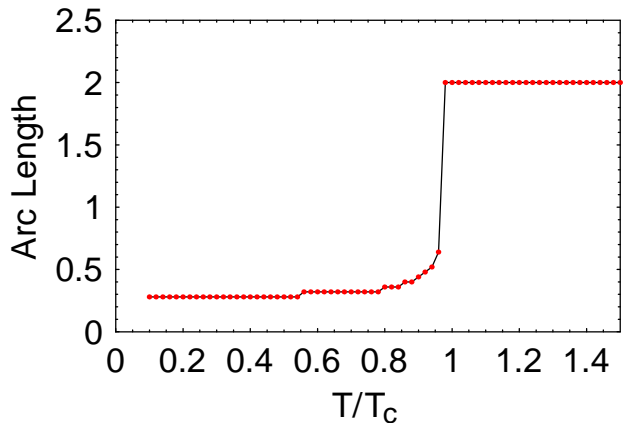}
\end{center}
\caption{(Color online) Length of ``Fermi arc" defined in text at
$T=0$ in superconducting state vs.\ temperature $T/T_c$.  Elastic
scattering rate parameter $\Gamma_{el,N}=\Delta_0$ }
\label{fig:arclength_T}
\end{figure}

Since the scattering rate is temperature and frequency dependent,
the criterion for the position of the spectral peaks changes as
$T$ is increased, resulting in a temperature-dependent change in
the ``arc length", as shown in Fig.~\ref{fig:arclength_T}.  This
is a very weak temperature dependence until quite close to $T_c$,
since it is driven by the $T$-dependence of the gap in the
theory~\cite{TDahm:2005b}, and above $T_c$ the arc length in the
current model is fixed because the gap goes to zero.  A theory
including a pseudogap in the normal metallic dispersion would give
rise to a $T-dependent$ (linear-$T$ for a marginal Fermi
liquid-like scattering rate\cite{CMVarma:1989}) arc length
\cite{JGStorey:2007,AVChubukov:2007,MRNorman:2007}.  Here we have
shown that, at low $T$ in the superconducting state, Fermi arcs
are also possible, but only in a dirty system.  Thus the recent
observation by Kanigel et al.~\cite{Kanigel:2007} of the collapse
of the arc length at low $T$ in the superconducting state places
constraints on the elastic scattering rate.  Since no arc is
observed at low $T$ in optimally doped samples, the scattering
rate must be a tiny fraction of $\Delta_0$, inconsistent with
ARPES determinations of such rates at low $T$, but consistent with
the STS result\cite{JAlldredge:2007} of $\Gamma (\omega \simeq
\Delta_0) \sim$ 2 meV near optimal doping. On the other hand, STS
has shown that the linear fits to the scattering rate yield a
tenfold increase in this rate for highly underdoped samples. In
this case, $\Gamma$ becomes an appreciable fraction of $\Delta_0$
for subgap energies.  The implication is therefore clear that
ARPES should eventually observe arcs in the superconducting state
as the system is underdoped.

Finally, it is worth observing that in Fig.~\ref{fig:spectral}
(a),(c) the EDC width at the antinodal points is significantly
larger than at the antinode  in the clean case shown
in~\ref{fig:spectral}(a)), despite the fact that the antinodal
scattering rate does not exceed the nodal one until energies
several times $\Delta_0$.  This is primarily because the EDC's are
dominated by quasiparticles ``on the mass shell"
$\omega=\Delta_{\bf k}$.  Thus, while the nodal scattering rate is
actually larger than the antinodal scattering rate for
$\omega=\Delta_0$, as shown in Fig.~\ref{fig:combined}, the
scattering rates which broaden $A({\bf k},\omega)$ are associated with
on-shell $\omega=\Delta_k$ energies. In this case, the antinodal
scattering rate refers to quasiparticles of energy
$\omega\sim\Delta_0$, while the nodal scattering rate to those of
energy $\omega\sim0$.  In addition, if interactions are
sufficiently strong, the inelastic scattering at the antinode will
eventually contribute, leading to a pronounced asymmetry of the
spectral function in the form of a large tail on the high binding
energy side in ARPES.  This may contribute significantly to
measured EDC widths if electron-electron scattering is
sufficiently strong.

~
\vskip .2cm

\section{Effect of sample inhomogeneity}
\label{sec:inhom}

\subsection{Low energy homogeneity}  Thus far we have considered a
homogeneous $d$-wave superconductor with inelastic scattering and
microscopic  disorder, which we treated by averaging to obtain a
translationally invariant self-energy characterized by momentum
${\bf k}$.  We saw in the previous section that at a given
quasiparticle energy, calculating a ``local self-energy" $\sum_{\bf k}
\Sigma({\bf k},\omega)$ provided a reasonable description of the LDOS
broadening. This assumes, as discussed in the introduction, that
the system is self-averaging within a region with a single gap
magnitude $\Delta_{\bf k}$. However, the BSCCO-2212 system is known to
be inhomogeneous at the nanoscale, in a somewhat mysterious way.
Spectra measured near the typical gap edge of a given sample are
extremely inhomogeneous, with coherence peak positions varying up
to a factor of 2-3 within a typical experimental field of view.
On the other hand, below some typical energy of order 30 meV,
near-homogeneity is recovered~\cite{KMLang:2002}.

Let us consider the
STM-determined low-energy local scattering rates.  These are quite
small relative to rates determined by ARPES, but increase with
increasing energy.  It is our hypothesis that quasiparticles at
low energy scatter so seldom that they explore many gap
``patches", and thus do {\it not} carry knowledge of the local gap
value when collected by the STM tip.  As the energy increases,
eventually their mean free path drops until they are localized
within a single patch.  Conductance spectra taken at these higher
energies will therefore reflect the local gap at the position of
the STM tip. It is significant (a) that the energy where
inhomogeneity appears is of order $\Delta_0$ in Bi-2212,
 and b)
that the doping dependence of this energy scale is rather weak.

To calculate the energy-dependent mean free path of our system,
we will need  the velocity
of a typical Bogoliubov quasiparticle $\overline{\bf v}_{qp}$ at
different points near the Fermi surface at $T=0$.  The group
velocity for
a quasiparticle of momentum ${\bf k}$ in a clean BCS superconductor is $v_{\bf k}\equiv
 \nabla
 E_{\bf k}$.  Here however we are accounting for scattering of these
 quasiparticles by impurities and collisions; therefore there is an effective spread
 in momentum of a typical quasiparticle due to its lifetime
 broadening.  We therefore define the
 speed of a typical quasiparticle near the Fermi surface with momentum
 ${\bf k}=(k_\perp,k\parallel)$ and energy at the gap edge
 to be

 \bea
\overline{ v}_{k_\parallel}= \sqrt{\sum_{k_\perp} v_{\bf k}^2
A({\bf k},\omega=\Delta_{\bf k})}\label{vqp}
 \eea

\begin{figure}
\begin{center}
\includegraphics[width=.45\columnwidth]{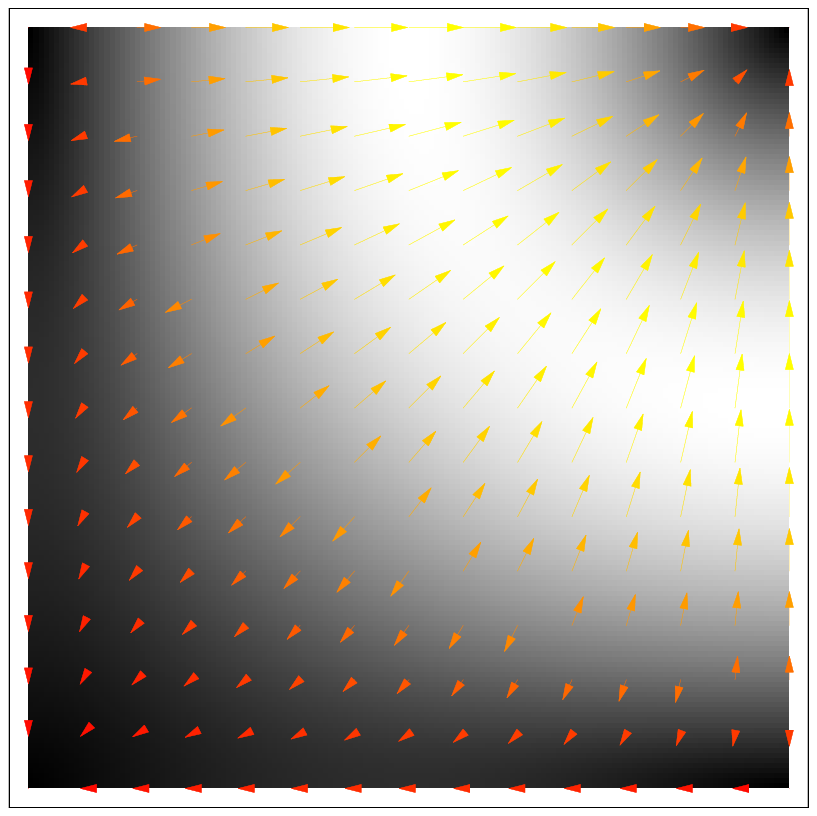}
\includegraphics[width=.45\columnwidth]{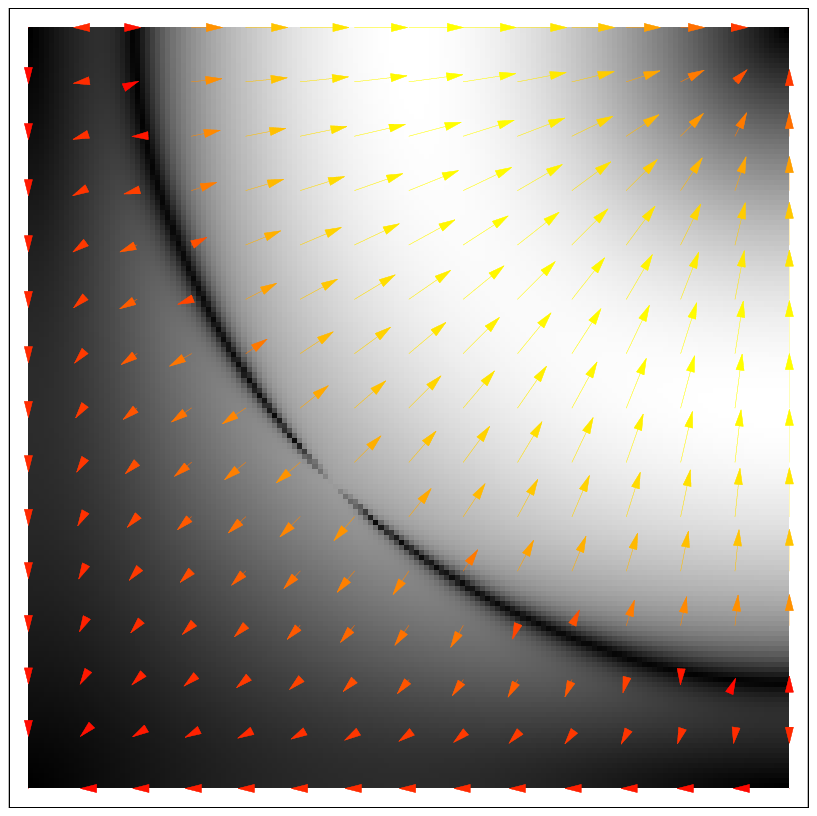}
\end{center}
\caption{(Color online) Quasiparticle group velocity for band
structure given in (\ref{band}) for a) normal state; b)
superconducting state.  Length of arrows or grayscale indicate
size of local quasiparticle speed. } \label{fig:broadening}
\end{figure}~

At the node, the spectral function is (neglecting real parts of
self-energies) $A({\bf k}_N,\omega) \simeq
(\Gamma/\pi)/(\xi^2+\Gamma^2)$, so using (\ref{vqp}) we find
${\overline v}_{qp}({\bf k}_N) =v_F({\bf k}_N)$.   At the antinode, the
spectral function is approximately\bea
 A({\bf k}_A,\omega)\simeq\frac{\Gamma}{\pi}\frac{2\Delta_{0}\left(\Delta_{0}+\xi\right)}{\xi^{4}+4\Delta_{0}^{2}\Gamma^{2}},
\eea leading to\bea {\overline
v}_{qp}({\bf k}_{A})\approx\sqrt{\frac{2\Gamma}{\Delta_{0}}}\,\,v_{F}({\bf k}_{A})\label{vqp_A},
\eea where we have assumed $\Gamma\ll \Delta_0$.  Thus the
``typical" antinodal velocity is generally much smaller than the
normal state Fermi velocity.  Using the STM determined value of
$\Gamma=2 $ meV at optimal doping, $\bar v_{qp}$ should be
suppressed by a factor of 3-4 relative to the Fermi velocity. This
is confirmed by a full evaluation of (\ref{vqp}) using the self
energy given by Eq.~\ref{total}.

\begin{figure}
\begin{center}
\includegraphics[width=.49\columnwidth]{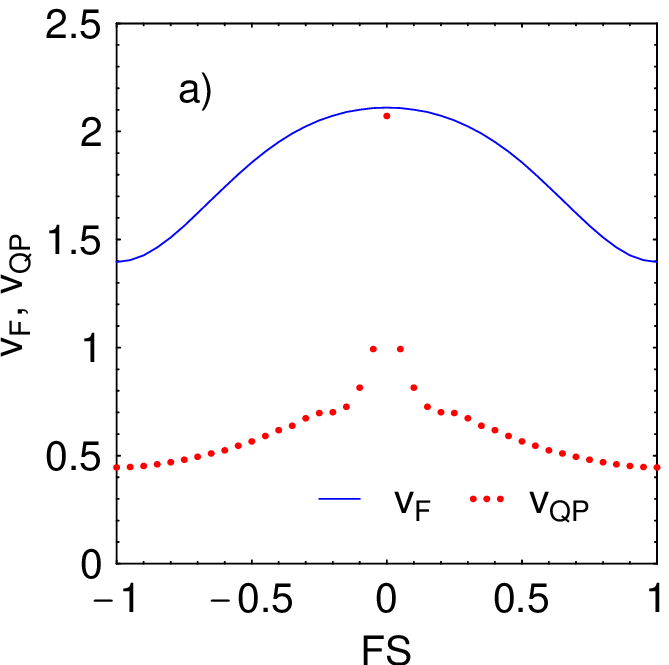}
\includegraphics[width=.49\columnwidth]{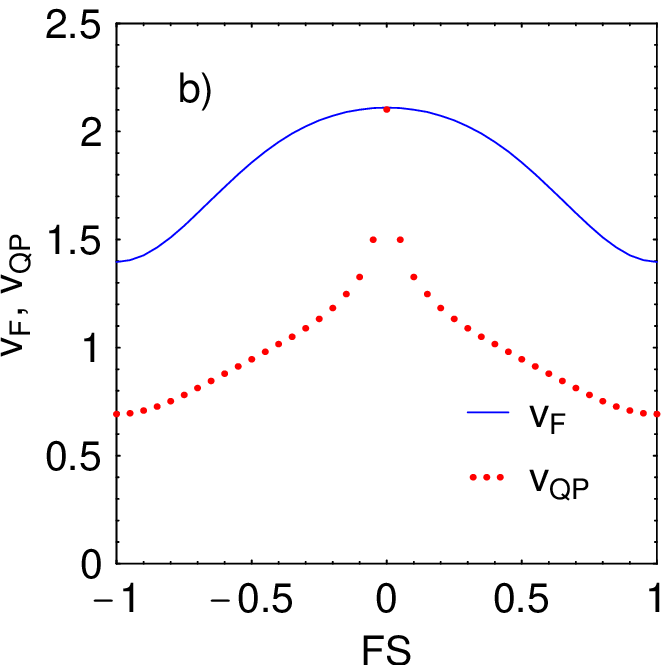}
\end{center}
\caption{(Color online) Typical quasiparticle group velocity for
band structure given in (\ref{band}) vs.\ position on Fermi
surface using expression (\ref{vqp}). The inelastic scattering is
calculated for $\bar U = 2.36 t$ while the elastic scattering
corresponds to $\Gamma_N (T_c)=0.05\Delta_0$ (a) and $\Gamma=0.2
\Delta_0$ (b).} \label{fig:groupvel}
\end{figure}~

\begin{figure}[t]
\begin{center}
\includegraphics[width=.49\columnwidth]{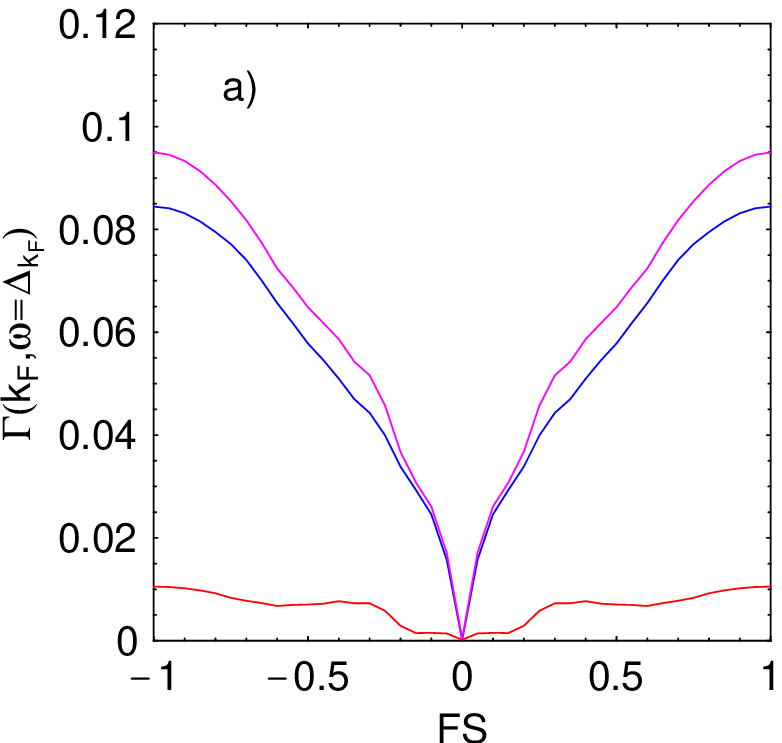}
\includegraphics[width=.49\columnwidth]{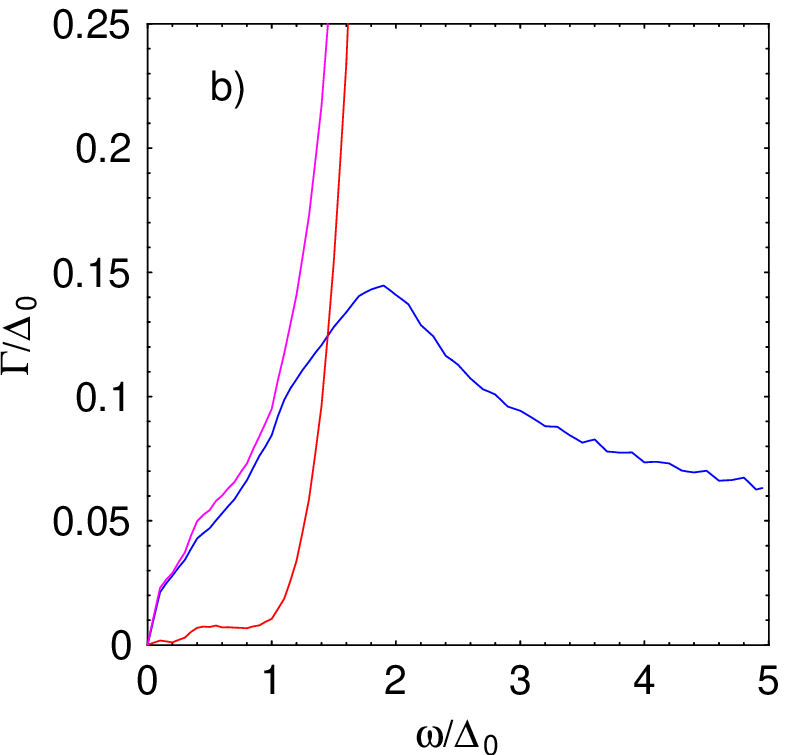}
\includegraphics[width=.49\columnwidth]{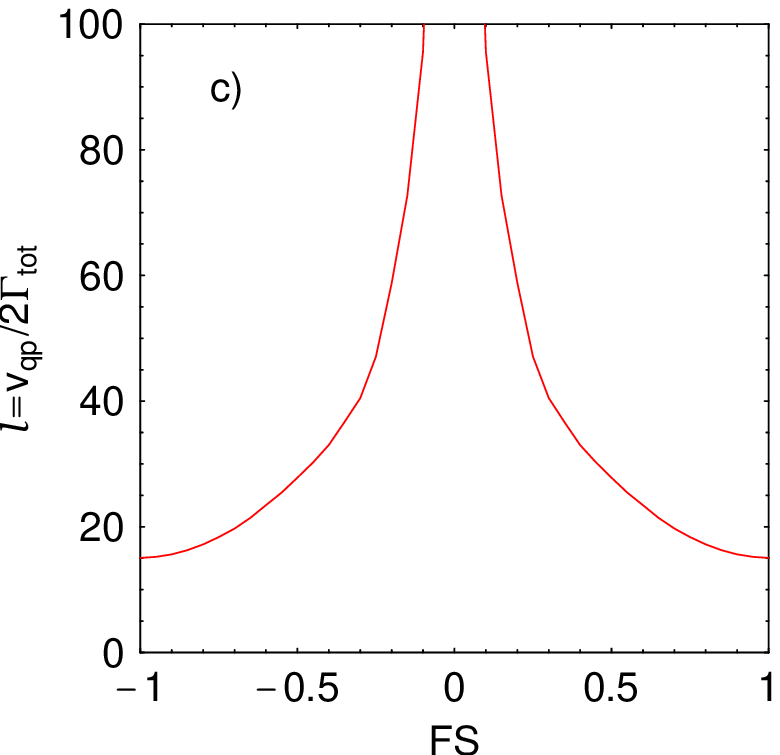}
\includegraphics[width=.49\columnwidth]{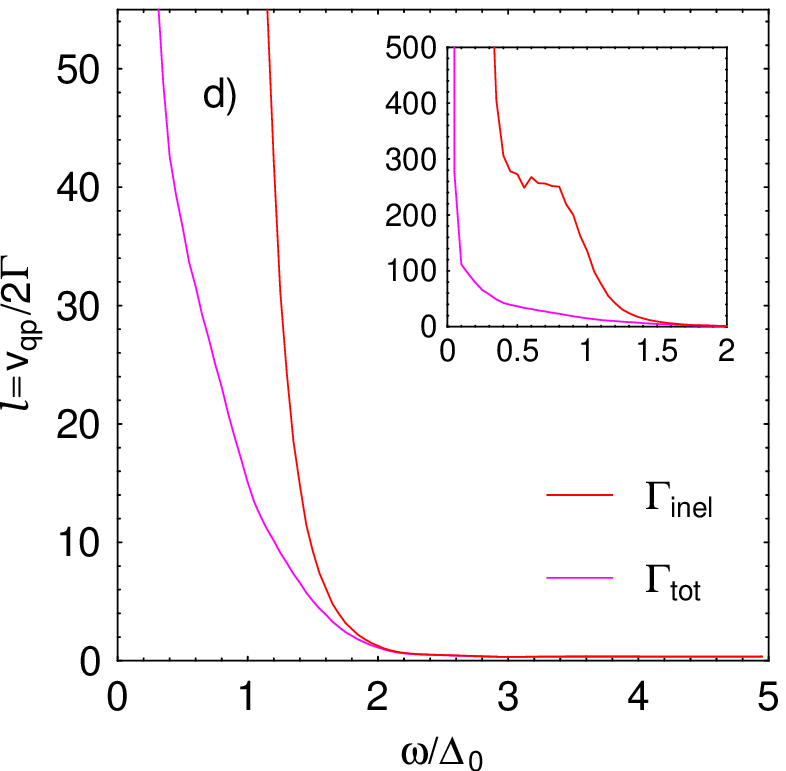}
\end{center}
\caption{(Color online) (a) on-shell scattering rate $\Gamma_{\rm
tot}({\bf k},\omega=\Delta_{\bf k})$ as a function of ${\bf k}$ from the node to
the antinode along the Fermi surface parameterized from -1 to 1,
for scattering parameters  elastic scattering parameters
$\kappa=1$, $\Gamma_{\rm el}(T_c)=0.1\Delta_0$ and $\bar U=2.36t$.
Blue: elastic scattering; red: inelastic; magenta: total
scattering rate.  (b) same quantity plotted vs.\ $\omega$.  (c,d)
Mean free path $\ell\equiv v_{qp}/(2\Gamma_{\rm
tot}({\bf k},\omega=\Delta_{\bf k})) $ vs.\ ${\bf k}$ and $\omega$ as in (a,b). }
\label{fig:onshell}
\end{figure}

We now define the on-shell quasiparticle mean free path \be
\ell_{\bf k}\equiv {\bar v_{qp}\over 2\Gamma(\omega=\Delta_{\bf k})},\ee
where $\Gamma$ may represent the elastic, inelastic, or total
one-particle scattering rate.  For extremely dirty systems,
momentum is not a good quantum number, and the on-shell quantities
may not provide a good representation of the mean free path.
However for the doped cuprate samples of interest it appears to us
that this is a reasonable estimate.  The on-shell scattering rate
is precisely that defined for the $\omega-$dependent momenta ${\bf k}$
corresponding to the points of the contours of constant
quasiparticle energy $E_{\bf k}$ which lie on the Fermi surface, i.e.
the so-called ``banana tips" defined within the octet
model\cite{JEHoffman:2002}. In Fig.~\ref{fig:onshell}, we plot
both the scattering rates and mean free paths defined in this way.
These are shown both as a function of $\omega$ and, equivalently,
of ${\bf k}$, for the set of parameters corresponding to a clean system
with inelastic scattering rate parameter $U=2.36t$ consistent with
neutron data, as in Fig.~\ref{fig:combined} (a) and (b) and
Fig.~\ref{fig:LDOS} (a) and (b).  In Fig.~\ref{fig:onshell} (a)
and (b), we see that the effective on-shell scattering rate is
indeed quasi-linear in energy over the range $0<\omega \lesssim
\Delta_0$, resulting in a momentum dependence similar to the
$d$-wave gap itself. The parameters chosen to approximately
reproduce the neutron scattering behavior in YBCO ( Fig.
\ref{fig:combined} (a)-(b)) appear to give a scattering rate scale
of approximately 0.1$\Delta_0$,  of order  the correct scale of
2meV determined by STS for BSCCO.

The mean free path in Fig.~\ref{fig:onshell} is seen to fall
rapidly with increasing energy, and becomes of order the patch
size $\sim 10a$ at an energy around the gap edge. We will
designate this energy $\omega_{loc}$ since it is only above this
energy where true local behavior characteristic of a particular
gap patch order parameter is measured. This is reminiscent of the
behavior identified by STS on BSCCO-2212 samples: below a critical
energy $\omega_{loc}$ of order or slightly less than $\Delta_0$,
spectra are homogeneous.  The famous inhomogeneity measured by STS
in this system appears only above this energy, and in particular
at and around the average gap energy (and above for underdoped
samples).  The energy $\omega_{loc}$ above which the sample
inhomogeneity is sensed by the STS measurement is therefore where
spectra measured in different locations begin to differ from one
another, and must therefore necessarily be associated with a
change in slope $N(\omega)$ at $\omega_{loc}$.  Such a kink in
typical spectra where homogeneity is lost has indeed been observed
in STS data~\cite{KMLang:2002,KMcElroy:2005}.

We note also the relative insensitivity of the observed kink
energy to doping $p$\cite{QPI_extinct}, despite rapid changes in
both the scattering rate and gap size over the measured range of
$p\simeq 0.08-0.2$. Within our picture, this can be understood at
least in part by the increase of the antinodal typical
quasiparticle velocity with the scattering rate as shown in
Eq.~(\ref{vqp_A}). Changes in Fermi surface shape may also play a
role in keeping the mean free path at this energy roughly doping
independent.

Finally, we comment on the significance of these results for the
quasiparticle interference (QPI) patterns observed in Fourier
transform STS on these systems\cite{JEHoffman:2002}.  It is
important to recall that the QPI patterms arise due to the random
static potential from  impurities. Static disorder can enhance
noise which may then swamp certain Fourier transform $\bf
q$-peaks\cite{LZhu:2003}, but it cannot suppress or broaden those
peaks in ${\bf q}$-space.  Inelastic scattering however will dephase
quasiparticle wavefunctions, and it is to be expected that a given
FT-STS ${\bf q}$-peak  will be lost when the inelastic scattering rate
becomes large enough such that the de Broglie wavelength of a
quasiparticle at the Fermi level is smeared by a significant
fraction.  The theoretical question of how such quasiparticle
interference patterns are destroyed is an interesting one, which
we are currently studying in more detail.  At present, however,
our results suggest that a qualitative increase in this smearing
occurs in the neighborhood of $\omega^*\sim\Delta_0$ due to the
rapid rise of the inelastic scattering.  It is {\it not} clear
from our analysis that the two energy scales $\omega_{loc}$ where
the system inhomogeneity becomes manifest and the scale $\omega^*$
where QPI patterns are destroyed are the same scale, but they are
of the same order, and experimentally appear to be quite
close\cite{QPI_extinct}.

\vskip .2cm

\subsection{STS and ARPES quasiparticle relaxation rates}

The size of the ARPES laser spot on the sample surface is many
times larger than a typical ``gap patch" size in BSCCO-2212.  This
fact has been frequently pointed out, but its consequences for
extracted quasiparticle lifetimes has not been explored to our
knowledge. We noted in the introduction that ARPES-extracted
lifetimes for both nodal and antinodal quasiparticles are much
larger than those determined recently by
STS~\cite{JAlldredge:2007}.  This fact can be plausibly explained
by taking our current knowledge of STS lifetimes and averaging the
spectral function over the distribution of gaps $P(\Delta)$ found
near the BSCCO-2212 surface.  A similar analysis was performed by
Fauqu\'e et al.\cite{BFauque:2007} to explain the large width of
the neutron resonance peak in BSCCO-2212, assuming that the same
distribution characterized the bulk of the sample probed by
neutrons. The gap distribution in STS has been found to have a
form roughly equal to a Gaussian centered at the average gap
$\overline \Delta$ and having a width $\sigma$ approximately equal
to $0.2\overline\Delta$, more or less independent of
doping~\cite{JAlldredge:2007}.  In Alldredge et al., the STS
scattering rate at the gap energy was extracted as $\sim\alpha
\omega$, as described in the introduction.  The average $\alpha$
for an optimally doped sample was $\langle \alpha \rangle \simeq
0.04$, giving a scattering rate at
$\omega=\overline{\Delta}\simeq$ 40meV of $\sim$ 2meV, or
$0.05\overline{\Delta}$.

\begin{figure}
\begin{center}
\includegraphics[width=.95\columnwidth]{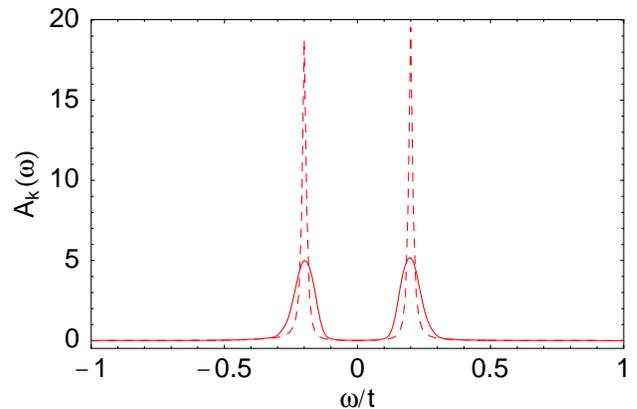}
\end{center}
\caption{(Color online) T=0 spectral function at antinode for
homogeneous $d$-wave superconductor with  scattering rate of
$\Gamma_{\rm el}(T_c)=0.05\Delta_0$ (dashed lines).  The solid
line is the same spectral function convolved with a Gaussian
distribution of gap values $P(\Delta_0)$ taken from
experiment~\cite{JAlldredge:2007}. Full width at half-maximum of
the solid curve is about 5$\times$ that of the dashed curve. Note
no instrumental energy broadening has been added to this curve.}
\label{fig:broadening}
\end{figure}~

  The spectral
 function at the antinode where this gap value is achieved is
 shown in Fig.~\ref{fig:broadening} as the dashed line, a narrow
 Lorentzian centered at $\Delta_0$.   But ARPES averages over many
 such lines, each centered at a different $\Delta_0$ and broadened
 by a different scattering rate, which depends on $\Delta_0$ in
 the above analysis.  The solid line in Fig.~\ref{fig:broadening}
 is then given by
 \bea
A_{\rm ARPES}({\bf k},\omega) \simeq \int P(\Delta_0)
A({\bf k},\omega;\Delta_0),\label{A_ARPES}
 \eea
where $A({\bf k},\omega;\Delta_0)$ is the spectral function calculated
in Section~\ref{sec:ARPES}.  Note to make this plot
 inelastic scattering was neglected in order to make the computation time
 practicable.  However our investigations showed that the distribution of
 center positions (gap values) is in any case much more important
 for the ultimate measured EDC width than the distribution of
 ($\Delta$-dependent)
 scattering rates.
This effective spectral function is roughly five times wider than
the ``intrinsic" spectral function which would be measured locally
in a small region where the local gap $\Delta(r)$ is
$\overline\Delta$.  Together with the instrumental resolution of
ARPES near the antinode of roughly 10meV, which should be
convolved with Eq.~(\ref{A_ARPES}), this gives an effective
antinodal EDC full width at half maximum of $\sim$ 25 meV.  This
is very similar to the full widths of antinodal ARPES EDC's
currently reported\cite{antinodeARPES}. It therefore appears to us
that the inhomogeneity effectively prevents ARPES (and planar
tunnelling measurements, which suffer from the same problem) from
measuring intrinsic lifetimes in the superconducting state of the
BSCCO family of materials.

\section{Conclusions}
\label{sec:conclusions}

  In this paper we have examined a series of questions arising from
recent STS measurements of lifetimes in BSCCO extracted from  fits
of the conductance to a BCS $d$-wave form with broadening,
Eqs.~(\ref{dwaveLDOS}) and (\ref{proportion}). We have argued that
an STS measurement probes local $d$-wave quasiparticle states
averaged over a mean free path, with the exception of unitary
bound states where quasiparticles are trapped on an atomic scale.
We have discussed different mechanisms for the scattering of
quasiparticles in the superconducting state, and presented the
results of model calculations within a BCS framework. Among these
scattering processes are elastic quasiforward weak scattering by
out of plane dopants, elastic pairing disorder scattering by the
same dopants, and elastic unitary isotropic scattering by in-plane
defects.  We also considered the effects of inelastic scattering
by spin fluctuations treated within a generalized random phase
approximation, for different values of the quasiparticle
interaction parameter, corresponding to a dynamical spin
susceptibility at $\pi,\pi$ with  a resonant mode contribution in
the $d$-wave state.

Within this model, the effective scattering rate rises with
energy, and is generally linear at low energy due to weak elastic
processes. If interactions are weak, quasiparticles over the whole
``low-energy" range $\omega \lesssim\Delta_0$ are scattered
primarily by impurities. If the system is sufficiently clean, or
if quasiparticle interactions are sufficiently strong such as to
induce a resonant mode, one may have a situation where the elastic
processes dominate only up to a subgap crossover energy $\lesssim
\Delta_0$. In the case with resonant mode considered in
Fig.~\ref{fig:inel_scattrate}(e)-(f), for example, this energy was
of order $0.5\Delta_0$ and the inelastic scattering rate for
near-nodal states rose rapidly above this energy; weaker
interactions move this crossover above the gap edge.
 The energy dependence of
the total scattering rate for any of these situations is never
strictly linear as a function of $\omega$. However, we found that
the model LDOS is well approximated by the imaginary part of a
$d$-wave Green's function with a linear self-energy.  This was
understood in terms of the effective on-shell total scattering
rate, which indeed appears to be quasilinear for reasonable
parameters.  Therefore, while the STS extraction of local
lifetimes is not particularly sensitive to the details of
scattering of the low energy states, it is roughly correct and
provides important insight regarding the doping dependence of
local scattering.

In this work we have focussed primarily on optimal doping, where a
BCS weak-coupling approach may be expected to work well. We have
therefore not addressed this striking doping dependence of the STS
scattering rate, where increases of a factor of ten or more
relative to optimal doping were reported in strongly underdoped
samples\cite{JAlldredge:2007}. Within the present theory, such
effects might be captured if the effective $\bar U/t$ were to
increase with underdoping, but this is difficult to describe in a
systematic manner in such a framework, as is well known.  In
addition, we have not addressed the measured correlation of the
scattering coefficient $\alpha$ with local gap size $\Delta_0$. In
fact, within the description of elastic potential scattering
discussed above, an anticorrelation would be obtained, in
contradiction to experiment. On the other hand, pair disorder
scattering, which we have also discussed briefly, appears to
provide a qualitatively correct description of this empirical
fact, as well as a quasilinear in $\omega$ local scattering rate.
While we have not provided a consistent microscopic description of
this physics, it appears to us to be worth exploring further.

As energy increases, the rising quasiparticle scattering rate
leads to a falling mean free path.  Our estimates show that at low
energy, quasiparticles explore (and self-average over) a large
area, providing an explanation of why STS spectra are homogeneous
at low energies.  At some critical energy $\omega_{loc}$ of order
$\Delta_0$, the mean free path becomes of order the gap patch
size, and STS  then (for $\omega >\omega_{loc}$) measures spectra
broadened by a local self-energy characteristic of a single gap.
The fact that a change in slope in conductance spectra is observed
at nearly the same energy in all patches appears to suggest that
the scattering rate rises fairly rapidly in this energy range.
Above this energy, conductance spectra are no longer homogeneous
since self-averaging over many gap patches no longer takes place.
It is furthermore to be expected--although we have not shown this
directly--that dispersive quasiparticle interference patterns
observed in Fourier transform STS will disappear at roughly this
energy as well, since their existence depends on the assumption of
a quasiparticle with definite energy scattering many times from a
disorder potential.  

To compare with ARPES, we have used the same model to calculate
the spectral function $A({\bf k},\omega)$, and shown that a ``Fermi
arc"---where the spectral peak remains at the Fermi level away
from the node in momentum space---may exist in the superconductin
state. We argued that while current experiments on optimally doped
samples do not see this feature, it should become visible in
underdoped samples as scattering rates increase. Actual  ARPES EDC
linewidths are much broader in energy than those  found within our
model of a homogeneous $d$-wave superconductor.   We have
therefore argued here that this is due in large part to shifts of
the spectral function caused by gap inhomogeneity, at least  in
the BSCCO family of cuprates, and estimated the effective EDC
width which should be measured in such a system.   Additional
effects which add to this large enhancement of EDC linewidths
include ARPES instrumental broadening and possible bilayer effects
near the antinode.   The current theory appears to account for the
much larger ARPES EDC widths in the superconducting state compared
to STS.

In this paper we have concentrated primarily on qualitative
physics, leaving open questions of quantitative fits to the ARPES
and STS spectra.
We hope to address these in a future publication.

\acknowledgments

The authors are grateful for discussions with  J.~Alldredge,
A.~Damascelli, J.C.~Davis,  D.~Maslov and P. W\"olfle. 
Work was begun at a
workshop supported by the Aspen Center for Physics, and was
partially funded by DOE Grant DE-FG02-05ER46236.  DJS would like
to acknowledge the Center for Nanophase Material Science at Oak
Ridge National Laboratory for support.

\section{Appendix 1: model calculation of scattering by pairing disorder}

The potential which one adds to the Hamiltonian is then (see Shnirman et al PRB 60,
   7517 (1999)).

   \bea
\hat V_1 = \sum_{{\bf k}{\bf k}'} (V_{\bf k} + V_{{\bf k}'})\tau_1,
   \eea
   where $V_{\bf k}=\delta\Delta \phi_{\bf k}$, and $\phi_{\bf k}$ is $(\cos k_x -\cos
   k_y)/2$, such that the average order parameter is $\Delta_0\phi_{\bf k}$.  The parameter $\delta\Delta$ has dimensions of energy
   and is is roughly the amplitude of the off-diagonal modulation.
   The disorder
   averaged self-energies then become (with label ``$\Delta$" for order
   parameter modulation
   scattering),

   \bea
\underline{\Sigma}^{\Delta}({\bf k},\omega) &=& n_i\sum_{{\bf k}'}
|V_{\bf k}+V_{{\bf k}'}|^2 \tau_1
\underline{G}({\bf k}',\omega) \tau_1 \\
\Sigma_0^{\Delta}({\bf k},\omega)&=& n_i(\delta\Delta)^2 \sum_{k'}
{\omega (\phi_{\bf k}^2
+\phi_{{\bf k}'}^2)\over \omega^2 - E_{{\bf k}'}^2} \\
\Sigma_1^{\Delta}({\bf k},\omega)&=& 2n_i(\delta\Delta)^2 \phi_{\bf k}
\sum_{k'}
{\phi_{{\bf k}'}\Delta_{{\bf k}'}\over \omega^2 - E_{{\bf k}'}^2}\\
\Sigma_3^{\Delta}({\bf k},\omega)&=& -n_i(\delta\Delta)^2 \sum_{k'}
{\xi_{{\bf k}'}(\phi_{\bf k}^2 +\phi_{{\bf k}'}^2)\over \omega^2 - E_{{\bf k}'}^2}\eea
   To gain some insight into the energy and momentum dependence of these quantities, we consider
   a model with a circular Fermi surface and approximate the  $d$-wave order parameter $\Delta_{\bf k}\simeq \Delta_\phi=\Delta_0 \cos 2\phi$.
   This approximation will not affect the qualitative
   low-energy dependence of the scattering rate.
   As before, we examine the pole of the Green's function to
   define the approximate total quasiparticle  scattering rate for a quasiparticle on the Fermi surface $\xi_{\bf k}=0$  as
   \bea
   \Gamma_{\rm el}^\Delta(\varphi,\omega) & = &
   -\mathrm{Im}\Sigma_{0}(\varphi,\omega)-\frac{\Delta_{\varphi}}{\omega}\mathrm{Im}\Sigma_{1}(\varphi,\omega)\nonumber
\eea and estimate the $\omega \ll \Delta_0$ self-energies at the
node $({\bf k}={\bf k}_N,\phi=\pi/4)$ and antinode $({\bf k}={\bf k}_A,\phi=0)$: \bea
\Gamma_{el,N}^\Delta\left(\omega\right)&\approx&
{1\over 4}\Gamma^\Delta\frac{\omega^{3}}{\Delta_{0}^{3}}\\
\Gamma_{el,A}^\Delta(\omega)&\approx& {5\over
4}\Gamma^\Delta\frac{\omega}{\Delta_{0}},\eea  and we have defined
$\Gamma^\Delta = \pi n_i N_0(\delta\Delta)^2$. It is clear that,
as intuitively expected, the scattering of quasiparticles by order
parameter modulations is largest near the antinodes, where the
mean order parameter is largest.  It is noteworthy that this
source of scattering is the only one considered here which gives
an anisotropic in ${\bf k}$ contribution on the Fermi surface as
$\omega\rightarrow 0$ which is largest at the antinode.
\begin{figure}[h]
\includegraphics[clip=true,width=.9\columnwidth,angle=0]{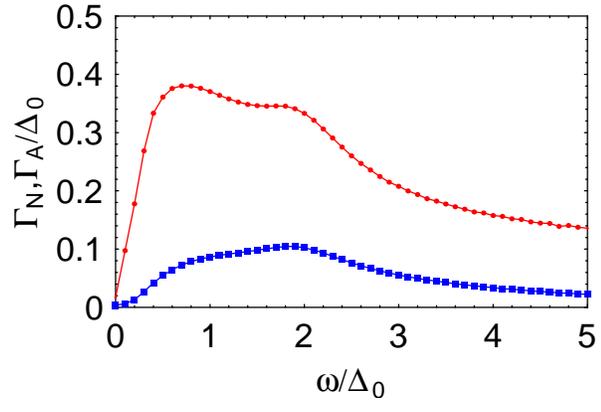}
\caption{Off-diagonal scattering rate at node (squares) and
antinode (circles) as function of energy $\omega$ for toy model
where order parameter modulation is confined to bonds neighboring
impurity site.} \label{fig:tau_1scatt}
\end{figure}


\begin{thebibliography}{99}

\bibitem{RHarris:2006}
R. Harris, P. J. Turner, Saeid Kamal, A. R. Hosseini, P. Dosanjh,
G. K. Mullins, J. S. Bobowski, C. P. Bidinosti, D. M. Broun,
Ruixing Liang, W. N. Hardy, and D. A. Bonn, Phys. Rev. B {\bf 74},
104508 (2006).

\bibitem{SOzcan:2006}
S. Ozcan, P. J. Turner, J. R. Waldram, R. J. Drost, P. H. Kes, and
D. M. Broun, Phys. Rev. B {\bf 73}, 064506 (2006).

\bibitem{TYamasaki:2006}T. Yamasaki, K. Yamazaki, A. Ino, M. Arita, H. Namatame, M. Taniguchi, A. Fujimori, Z.-X. Shen,
 M. Ishikado, S. Uchida, Phys. Rev. B {\bf 75}, 140513 (2007).

 \bibitem{antinodeARPES}  A. Damascelli, Z. Hussain, and Z.-X.
 Shen, Rev. Mod. Phys. {\bf 75}, 473 (2003).

\bibitem{JGStorey:2007} J.G. Storey, J.L. Tallon, G.V.M. Williams, J.W.
Loram, arXiv:0707.1549.

\bibitem{AVChubukov:2007} A. V. Chubukov, M. R. Norman, A. J. Millis, E. Abrahams,
arXiv:0709.1650.

\bibitem{MRNorman:2007} M. R. Norman, A. Kanigiel, M. Randeria, U. Chatterjee, J. C.
Campuzano, arXiv:0708.1713.

\bibitem{JAlldredge:2007} J.W. Alldredge, Jinho Lee, K. McElroy, M. Wang, K. Fujita, Y. Kohsaka, C.
Taylor, H. Eisaki, S. Uchida, P.J. Hirschfeld and J.C. Davis,
arXiv:0801.0087.

\bibitem{IMartin:2001}I. Martin and A.V. Balatsky, Physica C (Amsterdam) {\bf 357-
360}, 46 (2001).

\bibitem{ZWang:2002} Z. Wang, J. R. Engelbrecht, S. Wang,  H.
Ding, and S. H. Pan, Phys. Rev. B {\bf 65}, 064509(R) (2002).

\bibitem{SAKivelson:2003} S. A. Kivelson, I.P. Bindloss, E.  Fradkin,V Oganesyan, J.M.  Tranquada,   A. Kapitulnik,  and C. Howald,
Rev. Mod. Phys. {\bf 75}, 1201 (2003).

\bibitem{TSNunner:2005} T.S. Nunner, B.M. Andersen, A.
Melikyan and P.J. Hirschfeld,  Phys. Rev. Lett. {\bf 95}, 177003
(2005).

\bibitem{JXZhu:2006}J.-X. Zhu, arXiv:cond-mat/0508646.

\bibitem{ACFang:2006} A. C. Fang, L. Capriotti, D. J. Scalapino, S. A. Kivelson, N. Kaneko, M. Greven, and A. Kapitulnik,
Phys. Rev. Lett. {\bf 96}, 017007 (2006).

\bibitem{AVBalatsky:2006} A.V. Balatsky and J.-X. Zhu,  Phys. Rev.
B {\bf 74}, 094517 (2006).

\bibitem{EAbrahams:2000} E. Abrahams and C. M. Varma,
Proc. Natl. Acad. Sci. {\bf 97}, 5714 (2000).

\bibitem{TDahm:2005a}T. Dahm,  L.-Y. Zhu, P.J. Hirschfeld, and
 D. J. Scalapino,  Phys. Rev. B {\bf 71}, 212501 (2005).

 \bibitem{ZHS04} L.~Zhu, P.J.~Hirschfeld, and D.J.~Scalapino, Phys.~Rev.~B
{\bf 70}, 214503 (2004).

 \bibitem{TDahm:2005b} T. Dahm, P.J. Hirschfeld, D. J. Scalapino, and
L.-Y. Zhu,  Phys. Rev. B {\bf 72}, 214512 (2005).

\bibitem{BMAndersen:2006}B.M. Andersen, A. Melikyan, T.S. Nunner,
and P.J. Hirschfeld,  Phys. Rev. Lett.  {\bf 96}, 097004 (2006).

\bibitem{TSNunner:2006} T.S. Nunner,  W. Chen,  B.M. Andersen, A. Melikyan and
P.J. Hirschfeld, Phys. Rev. B {\bf 73}, 104511 (2006).

\bibitem{MMaska:2007}M. M. Maska, Z.Sledz, K. Czajka,
and M. Mierzejewski, Phys. Rev. Lett. {\bf 99}, 147006 (2007).

\bibitem{SHPan:2001} S. H. Pan {\it et al.},  Nature
{\bf 413}, 282 (2001).

\bibitem{KMLang:2002} K.M. Lang {\it et al.},  Nature {\bf 415}, 412 (2002).

\bibitem {CHowald:2001} C. Howald,  P. Fournier, and A. Kapitulnik,
Phys. Rev. B {\bf 64}, 100504(R) (2001).

\bibitem{MHHettler:1999} M.H. Hettler and P.J. Hirschfeld, Phys. Rev. B
{\bf 59}, 9606 (1999).

\bibitem{AShnirman:1999} A. Shnirman, I. Adagideli, P. Goldbart, and A.
Yazdani, Phys. Rev. B {\bf 60}, 7517 (1999).

\bibitem{QSB94}S.M.~Quinlan, D.J.~Scalapino, and N.~Bulut,
Phys.~Rev.~B {\bf 49}, 1470 (1994).

\bibitem{QHS96} S.M.~Quinlan, P.J.~Hirschfeld, and D.J.~Scalapino, Phys.~Rev.~B
{\bf 53}, 8575 (1996).

\bibitem{analytic} M.L.~Titov, A.G.~Yashenkin, and D.N.~Aristov,
Phys.~Rev.~B {\bf 52}, 10626 (1995).

\bibitem{JPaaske:2000} J. Paaske and D. Khveshchenko, Physica C 341-348 (2000)
265-266.

\bibitem{DSH01} D.~Duffy, P.J.~Hirschfeld, and D.J.~Scalapino,
Phys.~Rev.~B {\bf 64}, 224522 (2001).

\bibitem{earlycollectivemode} N. Bulut and D.J. Scalapino, Phys.
Rev. B {\bf 53}, 5149 (1996).

\bibitem{Eschrigreview}M. Eschrig, Adv. Phys. {\bf 55}, 47 (2006).

\bibitem{MRNorman:2007b}M.R. Norman, Phys. Rev. B {\bf 75}, 184514
(2007).

\bibitem{AAbanov:1999} A. Abanov and A. Chubukov, Phys. Rev.
Lett. {\bf 83}, 1652 (1999).

\bibitem{AAbanov:2002} A. Abanov, A.V. Chubukov, M. Eschrig, M. R. Norman, and J.
Schmalian, Phys. Rev. Lett. {\bf 89}, 177002 (2002).

\bibitem{MEschrig:2000}M. Eschrig and M.R. Norman, Phys. Rev.
Lett. {\bf 85}, 3261 (2000).

\bibitem{JFink:2006} J. Fink, A. Koitzsch,
 J. Geck, V. Zabolotnyy, M. Knupfer, B. Buchner, A. Chubukov,
 and H. Berger, Phys. Rev. B {\bf 74}, 165102 (2006).

\bibitem{IEremin:2005} I. Eremin, D. K. Morr, A. V. Chubukov, K. H.
Bennemann, and M. R. Norman Phys. Rev. Lett. {\bf 94}, 147001
(2005).

\bibitem{neutron_expt}S. Pailhes et al., Phys. Rev. Lett. {\bf 93}, 167001
(2004); S. M. Hayden et al., Nature (London) {\bf 429}, 531
(2004).

\bibitem{BFauque:2007} B. Fauqu\'e, Y. Sidis, L. Capogna, A. Ivanov, K. Hradil, C. Ulrich, A.I. Rykov, B. Keimer,
and P. Bourges, arXiv:cond-mat/0701052.

\bibitem{BWHoogenboom:2003}  B. W. Hoogenboom, C. Berthod, M. Peter,  \O. Fischer and A. A.
Kordyuk, Phys. Rev. B {\bf 67}, 224502 (2003).

\bibitem{GLdeCastro:2007}G.L. de Castro, C. Berthod, A. Piriou, E. Giannini, and \O.
Fischer, arXiv:cond-mat/0703131v2.

\bibitem{CMVarma:1989} C.M. Varma, P.B. Littlewood, S. Schmitt-Rink, E. Abrahams and A. E. Ruckenstein,
 Phys. Rev. Lett. {\bf 63}, 1996 (1989).

\bibitem{RSensarma:2007} Rajdeep Sensarma, Mohit Randeria, Nandini Trivedi,
Phys. Rev. Lett. {\bf 98}, 027004 (2007).

\bibitem{Kanigel:2007} A. Kanigel, U. Chatterjee, M. Randeria, M. R. Norman, S. Souma,
M. Shi, Z. Z. Li, H. Raffy, and J. C. Campuzano, Phys. Rev. Lett. {\bf 99}, 157001 (2007).

\bibitem{JEHoffman:2002} J.E. Hoffman,  K. McElroy, D.-H. Lee, K. M Lang, H. Eisaki, S. Uchida
and J.C. Davis, Science {\bf 297}, 1148 (2002).

\bibitem{KMcElroy:2005} K. McElroy, D.-H. Lee, J. E. Hoffman, K. M. Lang, J. Lee, E. W.
Hudson, H. Eisaki, S. Uchida, and J. C. Davis, Phys. Rev. Lett.
{\bf 94}, 197005 (2005).

\bibitem{QPI_extinct} Y. Kohsaka et al Submitted (to Nature).

\bibitem{LZhu:2003} Lingyin Zhu, W.A. Atkinson, and P. J.
Hirschfeld,   Phys. Rev. B {\bf 69}, 060503 (2004).

\end{thebibliography}
\end{document}